\documentclass[aps,prd,twocolumn,showpacs,floatfix,preprintnumbers,amsmath,amssymb,nofootinbib, superscriptaddress]{revtex4-1}

\usepackage{graphicx}
\usepackage{color}
\usepackage[font=small]{caption,subfig}
\usepackage{hyperref}

\renewcommand{\bm}[1]{\mathbf #1}

\newcommand{\lya}{{\ensuremath{{\rm Ly}\alpha}}}

\newcommand{\aap}{Astron. Astrophys.}
\newcommand{\apjs}{Astrophys. J. Suppl. Ser.}

\newcommand{\aj}{Astron. J.}

\newcommand{\mnras}{Mon. Not. R. Astron. Soc.}

\newcommand{\jcap}{Journal of Cosmology and Astroparticle Physics}
\newcommand{\pasj}{Publications of the Astronomical Society of Japan}

\begin{document}
\title{Stability of small-scale baryon perturbations during cosmological recombination}
\author{Tejaswi Venumadhav}
\affiliation{Theoretical Astrophysics Including Relativity (TAPIR), Caltech M/C 350-17, Pasadena, California 91125, USA}
\author{Christopher Hirata}
\affiliation{Center for Cosmology and Astroparticle Physics (CCAPP), The Ohio State University, 191 West Woodruff Lane, Columbus, Ohio 43210, USA}

\date{March 16, 2015}

\begin{abstract}
  In this paper, we study small-scale fluctuations (baryon pressure sound waves) in the baryon fluid during recombination. In particular, we look at their evolution in the presence of relative velocities between baryons and photons on large scales ($k \sim 10^{-1} \ {\rm Mpc}^{-1}$), which are naturally present during the era of decoupling. Previous work concluded that the fluctuations grow due to an instability of sound waves in a recombining plasma, but that the growth factor is small for typical cosmological models. These analyses model recombination in an inhomogenous universe as a perturbation to the parameters of the homogenous solution. We show that for relevant wavenumbers $k\gtrsim 10^3\ {\rm Mpc}^{-1}$ the dynamics are significantly altered by the transport of both ionizing continuum ($h\nu>13.6$ eV) and Lyman-$\alpha$ photons between crests and troughs of the density perturbations. We solve the radiative transfer of photons in both these frequency ranges and incorporate the results in a perturbed three-level atom model. We conclude that the instability persists at intermediate scales. We use the results to estimate a distribution of growth rates in $10^{7}$ random realizations of large-scale relative velocities. Our results indicate that there is no appreciable growth; out of these $10^7$ realizations, the maximum growth factor we find is less than $\approx 1.2$ at wavenumbers of $k \approx 10^{3} \ {\rm Mpc}^{-1}$. The instability's low growth factors are due to the relatively short duration of the recombination epoch during which the electrons and photons are coupled.
\end{abstract}

\maketitle
\section{Introduction}

The early universe is largely composed of atomic matter, or baryons, radiation and cold dark matter. The main resources available to study this era are the Cosmic Microwave Background (CMB) and Large Scale Structure (LSS). Primary anisotropies of the CMB are a result of the imprint of primordial fluctuations on radiation at early times \cite{PlanckInflation, Wmapcosmology}, while secondary anisotropies probe the matter distribution at late times \cite{PlanckLensing}. LSS surveys are a complementary probe of the clustering of matter at late times \cite{BOSS2013}.
 
The distribution and dynamics of baryons during early epochs of the Universe is poorly constrained by this data. The angular distribution of power in the CMB constrains them on large scales, through their coupling with the radiation and its effect on the Baryon Acoustic Oscillations. The CMB is well-described by a spectrum of adiabatic fluctuations at these scales -- these are motions of both the baryon and radiation fluids. Tight bounds exist on the primordial fluctuations of solely the baryon fluid at these scales -- the so-called isocurvature modes \cite{PlanckInflation}. 

This paper deals with the complementary limit of fluctuations in the baryon field on very small scales. In the CMB, this information is lost due to diffusion damping of the anisotropies. The spectral distortion associated with diffusion damping has been suggested as a probe of modes on these small scales \cite{Chluba2012}. The proposed PRISM mission aims to study CMB spectral distortions \cite{Prism2014}.

In the rest of this paper, we use the term ``matter" to refer to baryons, for reasons of readability; we are not concerned with the dynamics of cold dark matter. We concentrate on small-scale fluctuations of the matter field, and their evolution through the epoch of recombination. In particular, we undertake a detailed study of an instability which can amplify sub-Jeans length fluctuations at recombination suggested by Shaviv \cite{Shaviv1998}. The mechanism of interest is potentially applicable to wave numbers in the range $10^2\lesssim k\lesssim 3\times 10^5$ Mpc$^{-1}$ comoving. This is at much smaller scales than the standing acoustic waves responsible for peaks in the CMB power spectrum and baryon acoustic oscillations in the matter power spectrum (e.g. \cite{S65, EH98}), which are damped below the Silk scale \cite{Silk} $k_{\rm Silk} \sim 0.1$ Mpc$^{-1}$. We expect the pre-recombination amplitudes of modes at $k\gg k_{\rm Silk}$ to be extremely small, but if an instability is present then a ``seed'' amplitude could be generated by nonlinear generation of small-scale isocurvature modes \cite{PressVishniac}, or even thermal fluctuations if the growth rate is fast enough.

\begin{figure}[t]
  \includegraphics[width=\columnwidth]{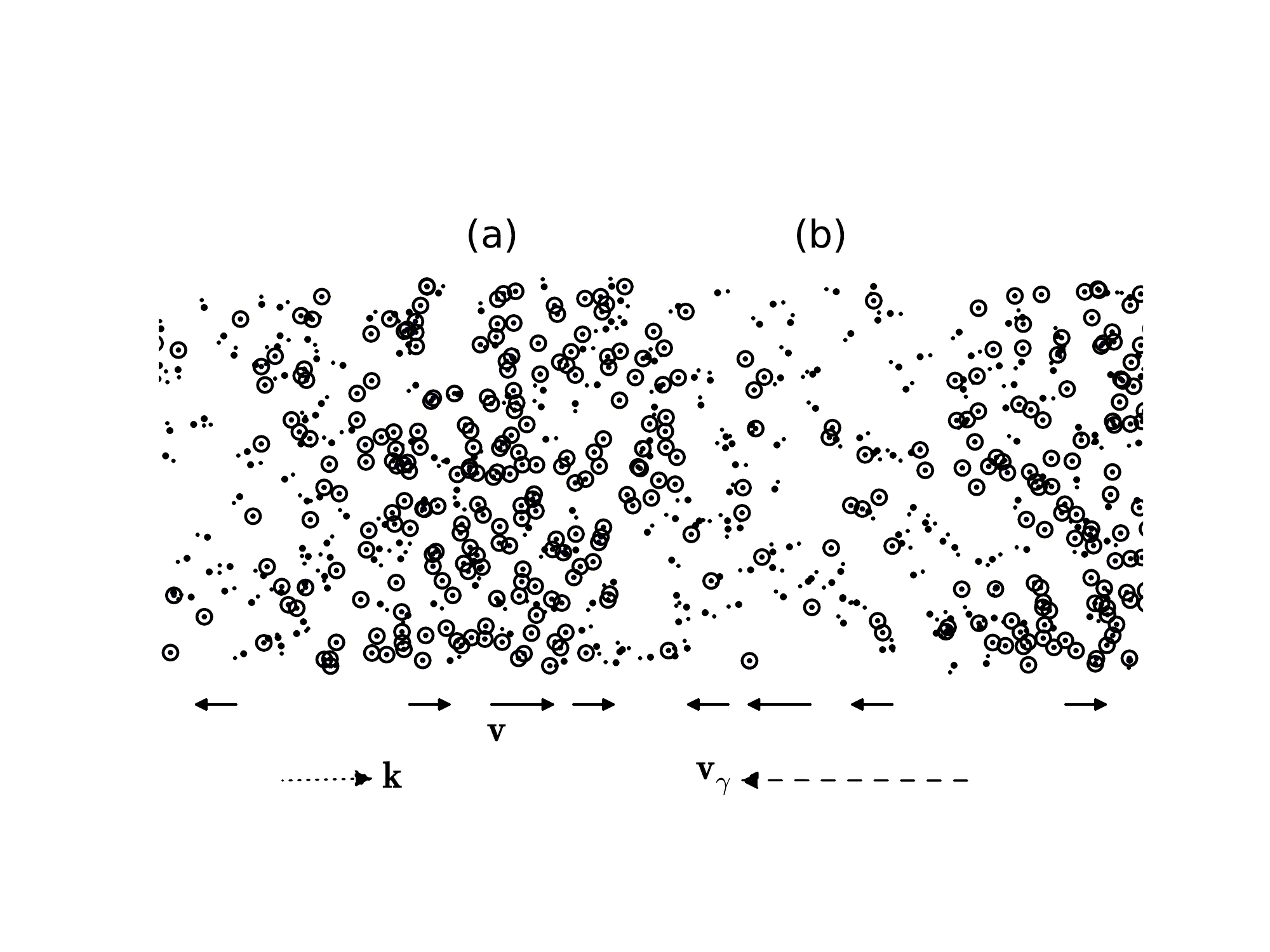}
  \caption{Illustration of the instability of sound waves during recombination: The symbol $\odot$ represents a neutral atom, while large and small dots represent positive ions and free electrons, respectively. The sound wave propagates to the right. Regions of compression and rarefaction, marked with (a) and (b), have lower and higher free electron fractions respectively. Solid arrows show the local velocity at various points along the wave in the bulk-rest frame of the matter. If the background flux of radiation, $\bm v_\gamma$, is directed to the left, the work done on the wave by the radiative force at (b) is larger than that extracted from it at (a).}
  \label{fig:cartoon}
\end{figure}

Shaviv's instability acts on sound waves propagating in a partially ionized gas, in the presence of a background flux of radiation. The scenario is illustrated in Fig.~\ref{fig:cartoon}. The key observation is that the fraction of ionized atoms is different in overdense and underdense regions; the ionization fraction, $x_{\rm e}$, is lower in overdense regions where recombination proceeds faster due to the increased flux of free electrons seen by the ionized atoms.

Sound waves are propagating longitudinal waves in the matter fluid -- if we orient ourselves along the wave-vector, $\bm k$, the local velocity at a compression is in the forward direction, while the opposite is true for rarefactions. Thus, the earlier observation leads to a negative correlation between the ionization fraction and the local velocity in the region of propagation.

In the presence of a background flux of radiation in the matter's bulk rest-frame, the radiative force acting on a mass element is related to the radiation flux, or alternatively its velocity $\bm v_\gamma$, by the opacity, which is proportional in turn  to the ionization fraction, $x_{\rm e}$. Over a time-period of the sound wave, the resulting force per unit mass $\bm a$ performs an amount of work $\Delta w$ given by
\begin{equation}
  \Delta w  = \oint \bm a \cdot d \bm r \sim \frac{u_\gamma \sigma}{m_{\rm H} c} \bm v_\gamma \cdot \oint x_{\rm e} \ d \bm r \mbox{,} \label{eq:workintegral}
\end{equation}
where in the second equation, the multiplicative factor involving the energy density of the radiation ($u_\gamma$), its interaction cross section with matter ($\sigma$), the particle mass (the hydrogen mass $m_H$) and the speed of light $c$ relates the force per unit mass to the ionization fraction. The net work done over a time period is nonzero due to the difference in ionization fractions during the forward and backward motion. From consideration of Fig.~\ref{fig:cartoon}, the work integral of Eq.~\eqref{eq:workintegral} is positive if the flux, $\bm v_\gamma$, is directed opposite to the wavevector, $\bm k$. 

The first estimate of the growth rates due to this mechanism, due to Shaviv \cite{Shaviv1998}, used the assumption of local thermal equilibrium (LTE) to derive the variations in the ionization fraction. Recombination in the real universe proceeds out of LTE, and most of the hydrogen first recombines to excited states before reaching the ground state \cite{Peebles, Zeldovich1969, Rybicki1994, HSSW, DS95, RecFast, HyRec, CosmoRec}. Subsequent work \cite{DiffusionTerm1, DiffusionTerm2} used the three level approximation to model non-LTE recombination, and incorporated the diffusion of microwave background photons, following which the expected growth rates were revised downward.

The standard treatment of recombination assumes that the ionization state is set by the local radiation field. This is valid in the homogenous case, since the transport of photons out of the region of interest is perfectly balanced by the influx from other regions. This is no longer true in the inhomogenous case, and these two components (the influx and outflux) do not balance each other. In particular, {\em direct} recombinations to the ground state, which did not affect the homogenous ionization fraction, $x_{\rm e}$, are important in determining its fluctuation, $\delta x_{\rm e}$. 

In this paper, we incorporate the transport of both continuum and Lyman-$\alpha$ (\lya) photons. We find simple analytical expressions for this ``non-local" contribution to the evolution of the ionization fraction, and provide revised estimates for the growth rates of the small-scale sound-waves.

The paper is organized as follows: In Section \ref{sec:informalderiv}, we expand upon the simple estimate given above for the work done on the fluctuations, and estimate the associated growth rates. In Section \ref{sec:background}, we list the relevant background variables, and the various factors which determine their size during the epochs of interest. 

In the subsequent sections, we write down equations of motion for the small-scale fluctuations. We start with the standard Newtonian equations for the density and velocity in Section \ref{sec:matterpert}. We estimate growth rates using a simple scaling relation for the ionization fraction fluctuation in Section \ref{sec:ionizationsaha}. We then move beyond this simple treatment, and study in detail the radiative transport of photons between different parts of the fluctuations -- Sections \ref{sec:continuum} and \ref{sec:Lya} deal with the transport of continuum and Lyman-$\alpha$ photons respectively. 

Finally, we bring all the pieces together and estimate the growth rates of the small-scale fluctuations in Section \ref{sec:solution}, and find their distribution due to a stochastic background of large-scale relative velocities in Section \ref{sec:average}. We finish with a short discussion of our results and their implications in Section \ref{sec:discussion}. We collect some details which lie outside the main line of analysis, but provide some physical intuition, into the appendices.

\section{Motivation and simple estimate} 
\label{sec:informalderiv}

This section closely follows the analysis of \cite{Shaviv1998}.

We use the two fluid approximation, where matter and radiation fluids are coupled by Thomson scattering of photons off free electrons. The characteristic response time, $\tau_{{\rm e}\gamma}$, is inversely related to the matter's opacity per unit mass, $\kappa$. For a given relative velocity between the two fluids, $\bm v_e - \bm v_\gamma = \bm v_{{\rm e}\gamma}$, the force per unit mass is expressed in terms of the response time as
\begin{equation}
\bm a = \frac{d\langle \bm v_{{\rm e}\gamma}\rangle}{d t} = \frac{\kappa}{c}{\bm F}_{\gamma} = -\frac{\langle \bm v_{{\rm e}\gamma}\rangle}{\tau_{{\rm e}\gamma}} \mbox{,} \label{eq:a0}
\end{equation}
where $\bm F_{\gamma}$ is the photon flux seen in the matter's rest frame. This force, and the related response time, are most easily obtained by considering the Doppler shifted background radiation field in the matter's rest frame. The result is \cite{Weymann1965}
\begin{equation} 
\frac{1}{\tau_{{\rm e}\gamma}} = \frac{4}{3} \frac{\sigma_{\rm T}}{m_{\rm H} c} a_{\rm rad} T_{\rm r}^4 x_{\rm e} \label{eq:opacity} \mbox{,}
\end{equation}
where $x_{\rm e}$ is the hydrogen ionization fraction, $\sigma_{\rm T}$ is the Thomson scattering cross-section and $a_{\rm rad}$ is the radiation energy density constant. The matter temperature, $T_{\rm m}$ closely follows the radiation temperature, $T_{\rm r}$, at these times. With this understanding, we omit the subscript on the temperature in subsequent equations.

Primordial adiabatic fluctuations entering the horizon lead to large-scale motions of the matter and radiation fluids. Their physical size, $\lambda_{\rm H}$ is $\approx 250 ~{\rm kpc}$ at recombination. Due to the small but finite response time, $\tau_{e\gamma}$, during this epoch, the matter velocity does not perfectly follow the local radiation velocity; this leads to a spectrum of relative velocities that can be estimated from the background cosmology \cite{Tseliakhovich2010}. 

We consider motions of the matter fluid alone, as contrasted with the large-scale adiabatic modes involving both matter and radiation. In particular, we concentrate the evolution of very small wavelength modes though the epoch of recombination out to late redshifts of $z=800$. We consider modes that are isothermal in nature, i.e., have a uniform matter temperature. As noted in the discussion (Section \ref{sec:discussion}), this condition restricts our analysis to modes with wavenumbers $k$ smaller than $\approx 3.5 \times 10^5\ {\rm Mpc}^{-1}$. The large scale adiabatic modes are effectively fixed on the timescales relevant to these small-scale modes, and provide a background radiation flux due to their associated relative velocity. The radiative force due to this flux is given by Eq.~\eqref{eq:a0}. 

The ionization fraction and opacity vary with the local density during recombination. Thus small-scale fluctuations of the matter density are associated with a modulation of the of the local force, denoted by $\delta \bm a$. The in-phase component of $\delta \bm a$ feeds power from the large-scale relative motions into small-scales.

The rest of this section estimates the size of this effect in a simplified scenario with direct recombination to the ground state of neutral hydrogen. With this assumption, the ionization fraction is given by the Saha equilibrium value, which we denote by $x_{\rm e}^{\rm S}$. This is set by the balance between the recombination of free electrons to the ground $1s$ state, and photoionization by microwave background photons.
\begin{equation}
\frac{(x_{\rm e}^{\rm S})^2}{1 - x_{\rm e}^{\rm S}} = \frac{(2 \pi m_{\rm e} k_{\rm B} T)^{3/2}}{h^3 n_{\rm H}} e^{-(E_{\rm I}/k_{\rm B} T)} \mbox{,} \label{eq:saha}
\end{equation}
where $E_{\rm I}$ is the ionization energy of a hydrogen atom in the ground $1s$ state, and $n_{\rm H}$ is the hydrogen number density. We take the logarithm of both sides of Eq.~\eqref{eq:saha}, and perturb it to estimate the power-law exponent relating the perturbed free electron fraction and hydrogen density as follows
\begin{equation}
  \alpha_{\rm S} = \frac{\delta \log{x_{\rm e}^{\rm S}}}{\delta \log{n_{\rm H}}} = -\frac{(1 - x_{\rm e}^{\rm S})}{(2 - x_{\rm e}^{\rm S})} \mbox{,} \label{eq:alpha}
\end{equation}
where we have used the assumption that the small-scale fluctuations do not perturb the temperature, $T$. The Saha electron fraction is approximately $x_{\rm e}^{\rm S} \approx 4 \times 10^{-3}$ at recombination, so the exponent $\alpha_{\rm S} \approx -0.5$. 

Consider a region with a background relative velocity between matter and radiation, $\bm v_{e,0} - \bm v_{\gamma,0} = \bm v_0$. The associated force per unit mass, $\bm a_0$, is related to the relative velocity $\bm v_0$ by the response time $\tau_{e\gamma}$, according to Eq.~\eqref{eq:a0}. The local matter density, velocity and force per unit mass are perturbed due to the small-scale fluctuation. For a sound wave, these perturbations are of the form
\begin{subequations}
\label{eq:pert}
\begin{align}
\frac{\delta\rho_{\rm m}}{\rho_{\rm m}} &= \delta_{\rm m} e^{i (\bm k \cdot \bm r - \omega t) } \mbox{,} \\
\bm v_{\rm m} &= v_{\rm {s, I} } \delta_{\rm m} \hat{\bm k} e^{i (\bm k \cdot \bm r - \omega t) } \mbox{,} \label{eq:vis} \\
\delta \bm a &= \frac{\delta\kappa}{\kappa} \bm a_0 = \frac{\delta x_e^{\rm S}}{x_e^{\rm S}} \bm a_0 \approx -|\alpha_{\rm S}| \delta_{\rm m} \bm a_0 e^{i (\bm k \cdot\bm r - \omega t) } \mbox{.} \label{eq:ais}
\end{align}
\end{subequations}
In the above relations, $v_{\rm{s,I}}$ denotes the isothermal sound speed. It is determined by the matter temperature according to $v_{\rm {s,I}} = [k_{\rm B} T(1 + x_{\rm e}^{\rm S})/m_{\rm H}]^{1/2}$. Averaged over the phase of the wave, the power input into the fluctuation by the extra force, $\delta \bm a$ of Eq.~\eqref{eq:ais}, is
\begin{align}
\langle p\rangle = \frac{1}{2}{\rm Re}(\bm v_{\rm m} \cdot \delta {\bm a}^\ast) &= -\frac{1}{2} |\alpha_{\rm S}| \delta_{\rm m}^2 v_{\rm{s,I}} \bm a_0 \cdot \hat{\bm k} \\
&= \frac{1}{2} |\alpha_{\rm S}| \delta_{\rm m}^2 v_{\rm{s,I}} \frac{\bm v_0 \cdot \hat{\bm k}}{\tau_{{\rm e}\gamma}} \mbox{.} \label{eq:inputp}
\end{align}
The first line uses Eq,~\eqref{eq:vis} and \eqref{eq:ais} for the velocity and force respectively, while the second uses Eq.~\eqref{eq:a0} for the background force and the definition of the response time $\tau_{{\rm e}\gamma}$ in Eq.~\eqref{eq:opacity}. The energy per unit mass in the fluctuation is $\langle\epsilon\rangle = (1/2) v_{\rm{m,max}}^2 = (1/2) \delta_{\rm m}^2 v_{\rm{s,I}}^2$. Hence the growth rate for the amplitude, $\mathcal{G}$, can be estimated from the input power of Eq.~\eqref{eq:inputp} as
\begin{equation}
  \mathcal{G} = \frac{\langle p\rangle}{2\langle \epsilon \rangle} = \frac{|\alpha_{\rm S}|}{2 \tau_{{\rm e}\gamma}} \frac{\bm v_0 \cdot \hat{\bm k}}{v_{\rm{s,I}}} \mbox{.}
\label{eq:growthrate}
\end{equation}
The growth of the instability is maximal during the epoch with large relative velocities and moderate response times. Relative velocities of the order of the isothermal sound speed are needed to produce an appreciable growth rate. The last part of Section \ref{sec:background} deals with the distribution of large scale relative velocities in detail. In particular, Figure \ref{fig:vrms} shows the mean relative speed, and the isothermal sound speed, as a function of redshift, $z$. We see that large relative velocities are much more probable in the post-recombination era; however, this effect is mitigated by the growing response time. Ultimately, the instability is limited by the relatively narrow duration of cosmic recombination.

\section{Background parameters}
\label{sec:background}

This section describes the relevant properties of the background on which the small fluctuations of interest live. 

We assume a standard spatially flat $\Lambda$ cold dark matter cosmology with the Planck cosmological parameters \cite{PlanckCosmology}. The derived quantities of interest to us are the hydrogen number density and ionization fraction, and the relative velocities between matter and radiation on large scales due to adiabatic fluctuations of primordial origin.

The simplest of these to obtain is the hydrogen number density, $n_{\rm H}$, which is given by
\begin{equation}
  n_{\rm H}(z) = 248.7 ~{\rm cm}^{-3} \left( \frac{1 + z}{1100} \right)^3 \frac{\Omega_b h^2}{0.022} \frac{1 - Y_{\rm He}}{0.752} \mbox{,}
\end{equation}
where $\Omega_b h^2$ is the Baryon fraction and $Y_{\rm He}$ is the Helium mass fraction.
 
It is considerably harder to estimate the hydrogen ionization fraction, $x_{\rm e}$, as a function of redshift. It is especially challenging to follow it through the epoch of recombination, when the universe transitions from a plasma of free electrons and hydrogen nuclei to a largely neutral phase with traces of free electrons that are strongly coupled to the cosmic microwave background (CMB) radiation. 

\begin{figure}[t]
  \includegraphics[width=\columnwidth]{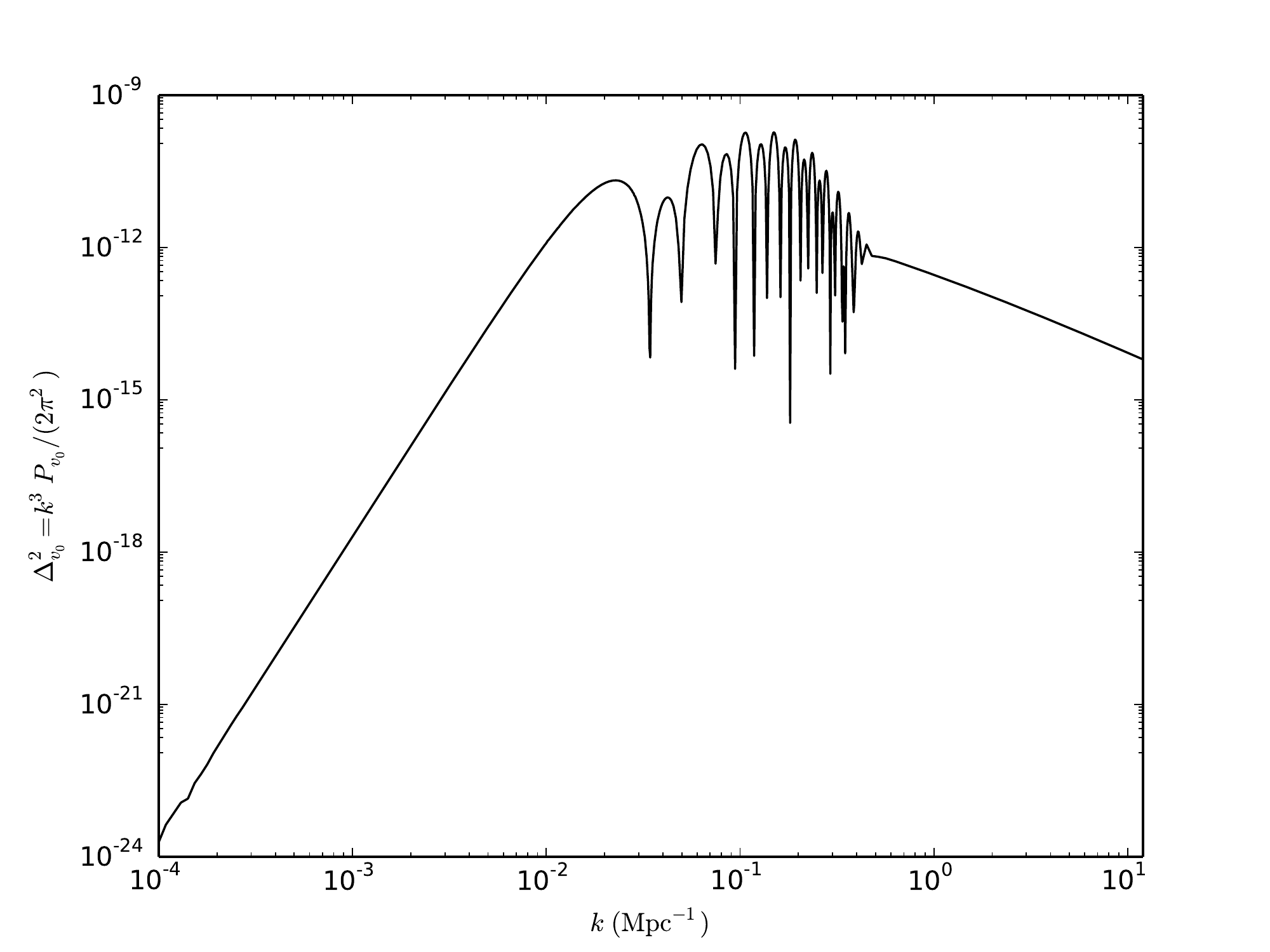}
  \caption{Power spectrum of relative velocities between matter and radiation at the redshift of recombination, $z_0 = 1100$. This assumes that these velocities arise from primordial adiabatic fluctuations. (This figure uses units in which velocity $\bm v_0$ is dimensionless).}
  \label{fig:pv0}
\end{figure}

This difficulty arises from the fact that direct transitions to the ground $1s$ state of hydrogen contribute very little to recombination, since they produce ionizing photons themselves. Instead, recombination mainly proceeds through excited states of neutral hydrogen. In order to derive the evolution of the ionization fraction to sub-percent level accuracy, we should follow the populations of a large number of excited states of the hydrogen atom \cite{HyRec, CosmoRec}.

We eschew this sophisticated analysis for a conceptually simpler, and less accurate, model of recombination originally proposed in Refs.~\cite{Peebles, Zeldovich1969}. This is adequate for the purposes of this paper, since we follow {\em fluctuations} in the ionization fraction. The errors introduced in the fluctuations by using the approximate model should be at the few-percent level.

This model approximates the hydrogen atom as a three level system; it assumes that the excited states of the true hydrogen atom are in thermal equilibrium with each other, and cascade down to the $n=2$ level through fast radiative decays. Atoms in the $2p$ state reach the ground state when photons redshift through the \lya\ line due to cosmological expansion, while those in the $2s$ level de-excite through a two-photon process. Direct recombination via the redshift of continuum photons is much slower (by a factor of $\sim 10^{-6}$) than through the \lya\ channel \cite{ContinuumEscape}. Hence we set the direct recombination's contribution to zero in the background case. As Section \ref{sec:continuum} demonstrates, this assumption is no longer valid in the perturbed case.

We add the recombination coefficients to the excited states to obtain an effective, or case B recombination coefficient, $\alpha_{\rm B}$. We also have an effective rate of photo-ionization from this state, $\beta_{\rm B}$. With these definitions, the ionization fraction evolves according to
\begin{equation}
  \dot{x}_{\rm e} = - C \Bigl( n_{\rm H} x_{\rm e}^2 \alpha_{\rm B} - 4 x_{1s} \beta_{\rm B} e^{-E_{21}/k_{\rm B} T} \Bigr) \mbox{,} \label{eq:peebles}
\end{equation}
where $C$ is the Peebles $C$-factor, which is the probability that an atom in the $n=2$ state reaches the ground state \cite{Peebles}. It is defined in terms of the \lya\ escape rate, the $2s$--$1s$ two-photon transition rate, and the rate of photo-ionization from the $n=2$ state. We derive explicit expressions for $C$ and the population of the $n=2$ state, $x_2$, in Section \ref{subsubsec:homogenous}.

The case B recombination coefficient and the effective photo-ionization rate are related by the principle of detailed balance \cite{Peebles, Zeldovich1969, RecFast}:
\begin{equation}
  \beta_{\rm B}(T) = \frac{(2 \pi m_{\rm e} k_{\rm B} T)^{3/2}}{4 h^3} e^{(E_2/k_{\rm B} T)} \alpha_{\rm B}(T) \mbox{.}
\end{equation}
We assume that the four sublevels of the $n=2$ level are equally occupied. Thus their occupation fractions are related by $x_{2p} = 3 x_{2s} = (3/4) x_{2}$. This is justified by the high effective $2p$--$2s$ transition rate at these redshifts ($\Lambda_{2p,2s} \approx 2.5 \times 10^4 ~{\rm s}^{-1}$ \cite{AH2010, HyRec}). This is much faster than both the case B recombination rate per hydrogen atom, and the photo-ionization rate, which are $\sim 1.3 \times 10^2 ~{\rm s}^{-1}$, and the two-photon decay rate, $\Lambda_{2s,1s}=8.22 ~{\rm s}^{-1}$ \cite{Goldman89}.

In deriving Eq.~\eqref{eq:peebles}, we assume that the population of the $n=2$ level is in steady state, i.e., we balance the net rate of recombination and photo-ionization against the escape of \lya\ photons and two-photon decays. This is valid if the abundance of intermediate states is very small; in this case, $x_2/x_{1s}$ can be estimated from the recombination codes themselves (e.g. Ref.~\cite{HyRec}), and is typically of order $\sim 10^{-14}$.

The final piece needed is the spectrum of relative velocities between matter and radiation on large scales. We assume that velocities are irrotational, i.e., they are aligned with their wave vectors, $\bm k$. The velocity at any point in space is a Gaussian random variable, whose two-point correlation function is 
\begin{align}
  \langle v_{0,i} (\bm x) v^\ast_{0,j} (\bm x) \rangle = \frac{1}{3} \delta_{i j} \int d \ln{k} \ \Delta^2_{v_0}(k) \mbox{,} \label{eq:distribution}
\end{align}
where $\Delta^2_{v_0}$ is the dimensionless power per log wave-number of the component along the wave-vector. This power spectrum is given by \cite{Ma1995}
\begin{align}
  \Delta^2_{v_0}(k) = \frac{k^3 P_{v_0}(k)}{2 \pi^2} = \frac{1}{k^2} \vert \Theta_{\rm m}(k) - \Theta_{\rm r}(k) \vert^2 \Delta^2_{\zeta}(k) \mbox{,}
\end{align}
where $\zeta$ is the primordial curvature perturbation, and $\Theta_{\rm m}$ and $\Theta_{\rm r}$ are the transfer functions for the matter and radiation velocity divergence respectively. We use the publicly available CLASS code to obtain these transfer functions \cite{classcode}. Figure \ref{fig:pv0} shows the resulting power spectrum for the relative velocity. We observe that most of the power is in scales near $k \sim 0.1 \ \rm{ Mpc}^{-1}$.

\begin{figure}[t]
  \includegraphics[width=\columnwidth]{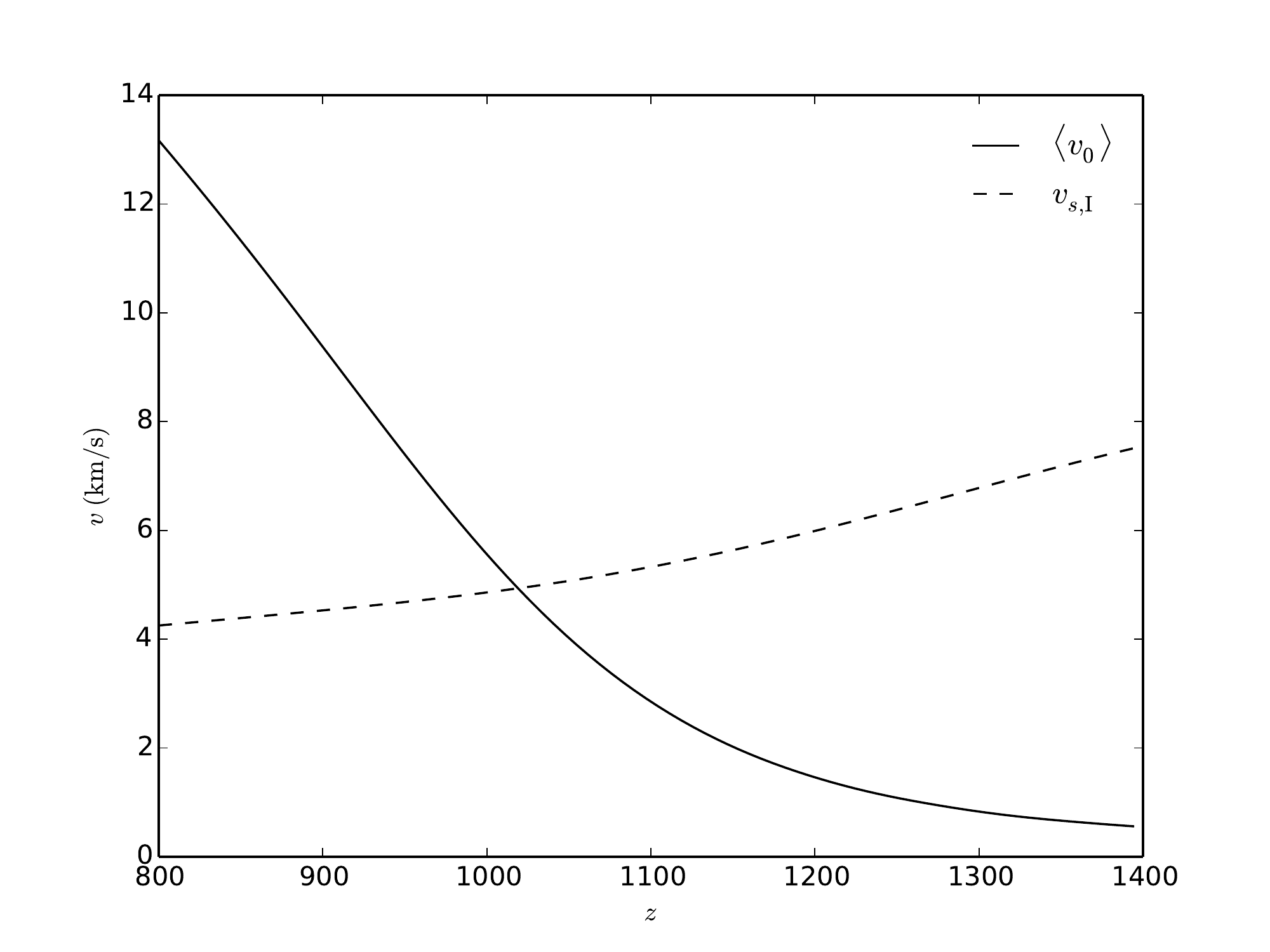}
  \caption{\label{fig:vrms} Speeds with redshift: the solid line shows the average magnitude of the background relative velocity between matter and radiation; the dashed line shows the isothermal sound speed.} 
\end{figure}

We estimate the typical velocities from the distribution of Eq.~\eqref{eq:distribution}. Figure \ref{fig:vrms} shows the both the average speed of the matter relative to the radiation, and the isothermal sound speed, as a function of redshift. We observe that these velocities are very small during the pre-recombination era: the matter-radiation response time, $\tau_{{\rm e}\gamma}$, is much smaller than the expansion age due to rapid scattering, which suppresses the relative velocities. During recombination the free electron fraction drops, and the response time becomes comparable to the expansion age, i.e., recombination leads to decoupling. 

\section{Linear analysis of density and velocity fluctuations}
\label{sec:matterpert}

Small-scale fluctuations of the matter field perturb the density, velocity and the ionization fraction. We denote the fractional matter overdensity by $\delta_{\rm m}$, the velocity by $\bm v_{\rm m}$ and the ionization fraction and its fluctuation by $x_{\rm e}$ and $\delta x_{\rm e}$ respectively. In addition to these, we denote the perturbed gravitational potential by $\delta \phi$. In this section, we derive the evolution equations for the density and velocity. In what follows, $\bm x$ is the position on a comoving grid, while a dot represents a derivative with respect to coordinate time.

There is a small amount of helium present in the early Universe: the He:H ratio by number, $f_{\rm He}$, is given in terms of the Helium mass fraction, $Y_{\rm He}$, by $f_{\rm He} = Y_{\rm He}/[4(1-Y_{\rm He})] \approx 0.08$. We consider late times, $z\lesssim 1800$, where the helium is fully neutral, so that it does not contribute to the ionization fraction. The hydrogen mass fraction $X_{\rm H} = 0.76$ is also used in the equations below.

The matter density, velocity, and gravitational potential on sub-horizon scales are governed by the Newtonian equations of motion -- the equation of continuity, the Navier-Stokes equation and Poisson's equation written in the comoving frame (as in \cite{KolbTurner}). The linearized forms of these equations are:
\begin{subequations}
\label{eq:newtonian2}
\begin{align}
  \dot{\delta}_{\rm m} + \frac{1}{a} {\bm \nabla} \cdot {\bm v_{\rm m}} & = 0 \mbox{,} \\
  \dot{\bm v}_{\rm m} + H {\bm v_{\rm m}} & = - \frac{1}{a \rho_{\rm m}} \bm \nabla P - \frac{1}{a} \bm \nabla \delta \phi + \bm f_{{\rm rad}} \mbox{,} \label{eq:forcebalance} \\
  \frac{1}{a^2} \nabla^2 \delta \phi &= - 4\pi G \rho_{\rm m} \delta_{\rm m} \mbox{.} 
\end{align}
\end{subequations}
The quantity $H$ is the Hubble rate of expansion, $H = \dot{a}/a$. The relative velocity force term, ${\bm f}_{{\rm rad}}$, depends on the flux of background radiation in the local matter rest frame. We use Eqs.~\eqref{eq:a0} and \eqref{eq:opacity} to write the force as $\bm f_{{\rm rad}} = - \Lambda_{\rm e \gamma} X_{\rm H} x_{\rm e} (\bm v_{\rm m} - \bm v_{\rm r})$, where $\Lambda_{\rm e \gamma}$ is the inverse of the response time in the case where the hydrogen is completely ionized and the helium mass is neglected. Typical large-scale relative velocities, $\bm v_0$, on comoving scales $k \approx 0.1~\rm{Mpc}^{-1}$, appear nearly uniform to the small-scale matter fluctuations. By definition, the latter do not perturb the radiation field. Hence the force associated with the relative velocity is
\begin{equation}
  \bm f_{{\rm rad}} = -\Lambda_{\rm e \gamma} X_{\rm H}x_{\rm e} \bm v_{\rm m} - \Lambda_{\rm e \gamma} X_{\rm H} \delta x_{\rm e} \bm v_0 \mbox{.} \label{eq:feqn}
\end{equation}
We decompose the velocity into scalar (curl-free) and vector (divergence-free) parts:
\begin{equation}
  \Theta_{\rm m} = \bm \nabla \cdot \bm v_{\rm m} {\rm ~~and~~}
  \Omega_{\rm m} = \bm \nabla \bm \times \bm v_{\rm m} \mbox{.}
\end{equation}
Under the equation of motion, \eqref{eq:forcebalance}, the vector part's evolution depends on the scalar part through the latter's modulation of the free electron fraction in the force term, but the reverse is not true. We focus on the scalar part in the rest of this paper.

We expand the restoring force due to the pressure up to first order in the fluctuation as follows
\begin{align} 
  - \frac{\bm \nabla P}{a \rho_{\rm m}}  &= -\frac{1}{a \rho_{\rm m}} \bm \nabla (n k_{\rm B} T) \nonumber \\
  &= -\frac{1}{a \rho_{\rm m}} k_{\rm B} T \bm \nabla [n_{\rm H} (1 + f_{\rm He} + x_{\rm e} + \delta x_{\rm e}) ] \nonumber \\
  &= -\frac{1}{a} \frac{k_{\rm B} T}{m_{\rm H}} X_{\rm H} \bm \nabla [ (1+ \delta_{\rm m}) (1+ f_{\rm He} + x_{\rm e} + \delta x_{\rm e} ) ] \nonumber \\
  &= -i \frac{\bm k}{a} \frac{k_{\rm B} T}{m_{\rm H}} X_{\rm H} [ (1 + f_{\rm He} + x_{\rm e}) \delta_{\rm m} + \delta x_{\rm e} ] \mbox{.} \label{eq:peqn}
\end{align}
We substitute the pressure and relative velocity force terms [Eqs.~\eqref{eq:peqn} and \eqref{eq:feqn}] in the Newtonian equations [Eq.~\eqref{eq:newtonian2}], and eliminate the gravitational potential, $\delta \phi$. Assuming plane-wave forms for the perturbed quantities, $\alpha(\bm x) = \int [d^3 {\bm k}/(2\pi)^3] \alpha(\bm k) \exp{(i \bm k \cdot \bm x)}$, the final forms of the evolution equations for the matter density and velocity are:
\begin{subequations}
\label{eq:newtonianfinal}
\begin{align}
  \dot{\delta}_{\rm m} &= - \frac{1}{a} \Theta_{\rm m} \mbox{,} \\
  \dot{\Theta}_{\rm m} &= - \frac{k^2}{a} \Bigl[ \frac{4\pi G \rho_{\rm m}}{k^2} a^2 - X_{\rm H}(1 + f_{\rm He} + x_{\rm e}) \frac{k_{\rm B} T}{m_{\rm H}} \Bigr] \delta_{\rm m} \nonumber \\
  & ~~~- ( H + \Lambda_{\rm e \gamma} X_{\rm H} x_{\rm e} ) \Theta_{\rm m} + X_{\rm H}\Bigl(\frac{k^2}{a} \frac{k_{\rm B} T}{m_{\rm H}} - i \Lambda_{\rm e \gamma} \bm k \cdot \bm v_0 \Bigr) \delta x_{\rm e} \mbox{.} \label{eq:deltav}
\end{align}
\end{subequations} 

\section{Ionization fraction fluctuation: Saha equilibrium scaling}
\label{sec:ionizationsaha}

In order to get a complete picture of the ionization fraction's evolution, we need to study the transport of photons between different parts of the fluctuations. Before we deal with this problem in Sections \ref{sec:continuum} and \ref{sec:Lya}, we make a simple first estimate following Ref.~\cite{Shaviv1998}.

The simplifying assumption in this section is that the ionization fraction scales with matter density in the same manner as the value calculated using local thermodynamic equilibrium (LTE, or Saha equilibrium). In subsequent sections, we consider non-equilibrium ionization. We note that perturbed non-equilibrium ionization in cosmology is one of the contributions to the CMB bispectrum and hence has been investigated as a potential contaminant to primordial non-Gaussianity studies \cite{Novosyadlyj, Khatri09, Senatore2009, Pitrou10} and probe of new physics \cite{Dvorkin13}, however these studies did not consider the very high $k$ of interest in this paper and hence did not have to solve the nonlocal radiative transfer problem considered in Sections \ref{sec:continuum} and \ref{sec:Lya}.

\begin{figure}[t]
\begin{center}
  \includegraphics[width=\columnwidth]{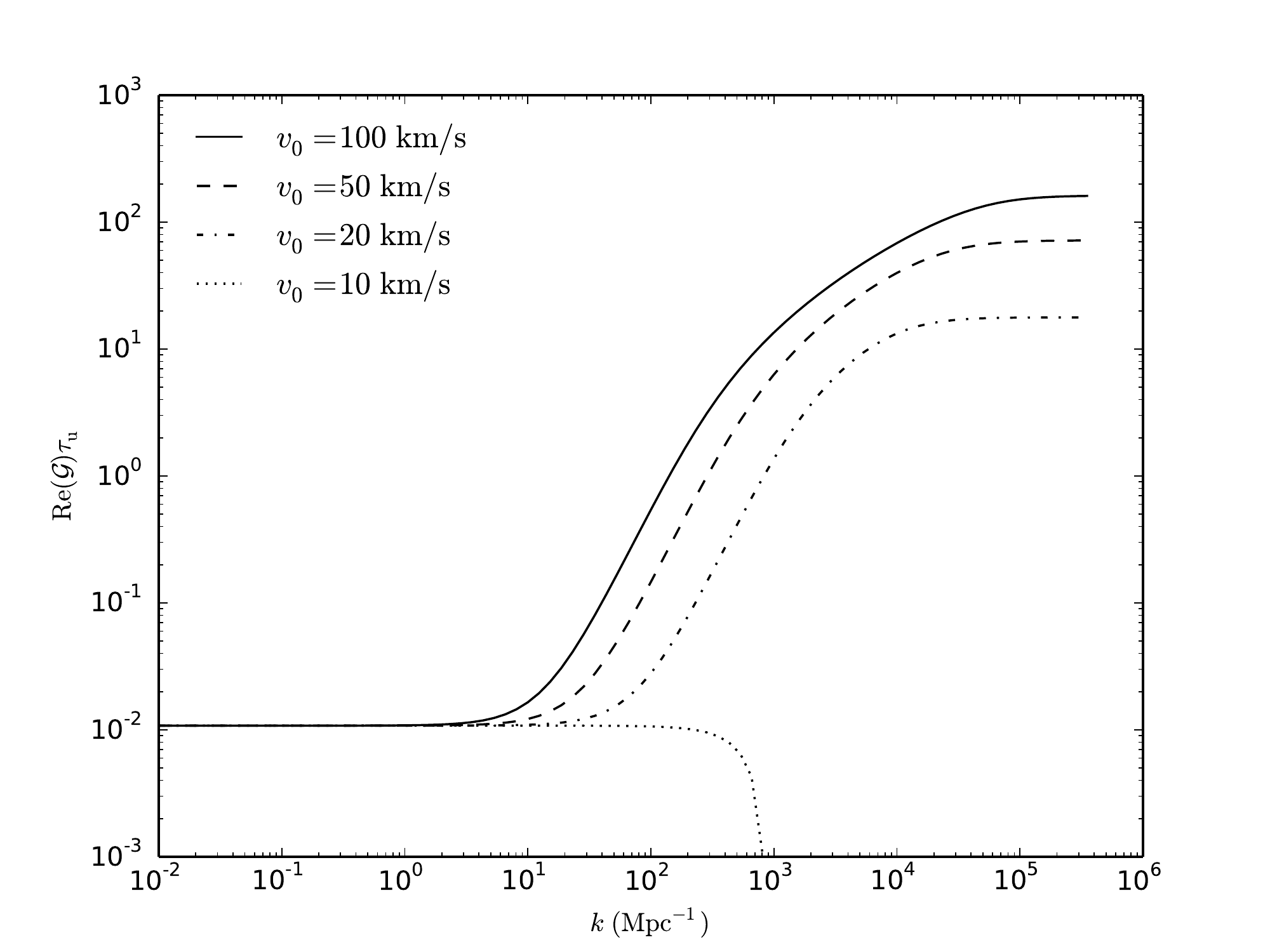}
  \caption{Maximum instantaneous growth rate for small-scale fluctuations in the matter field at recombination, normalized to the net elapsed coordinate time, $\tau_{\rm u}$. The wave-vector is oriented along the large-scale relative velocity between matter and radiation, $\bm v_0$. This approximates the perturbed ionization fraction with the scaling relation of the Saha equilibrium value, given by Eq.~\eqref{eq:alpha}.}
  \label{fig:growthsaha}
\end{center}
\end{figure}

Using the scaling of Eq.~\eqref{eq:alpha} for the ionization fraction fluctuation in Eq.~\eqref{eq:deltav}, we reduce the Newtonian evolution equations to
\begin{subequations}
\label{eq:deltavsaha}
\begin{align}
  \dot{\delta}_{\rm m} &= - \frac{1}{a} \Theta_{\rm m} \mbox{,} \\
  \dot{\Theta}_{\rm m} &= -\Bigl[ H + \Lambda_{\rm e \gamma} X_{\rm H} x_{\rm e} \Bigr] \Theta_{\rm m} + \biggl\{ \frac{(1 - x_{\rm e})}{(2 - x_{\rm e})} i \Lambda_{\rm e \gamma} X_{\rm H} x_{\rm e} \bm k \cdot \bm v_0 \nonumber \\
  & ~~~- \frac{k^2}{a} \Bigl[ \frac{4\pi G \rho_{\rm m}}{k^2} a^2 - \frac{2 - f_{\rm He} x_{\rm e}}{2 - x_{\rm e}} X_{\rm H} \frac{k_{\rm B} T}{m_{\rm H}} \Bigr] \biggr\} \delta_{\rm m} \mbox{.}
\end{align}
\end{subequations}
The instantaneous growth rate, $\mathcal{G}$, is the largest eigenvalue of the system of Eq.~\eqref{eq:deltavsaha}. Figure \ref{fig:growthsaha} plots this growth rate (normalized to a net elapsed coordinate time, $\tau_{\rm u}$, at the redshift of recombination, $z_0 = 1100$) for various values of the large-scale relative velocity, with the wave vector oriented along its direction. 
Modes with comoving wavenumbers satisfying $k > 2 \times 10^2 \ {\rm Mpc}^{-1}$ (or physical wavelength smaller than $\approx 30 \ {\rm pc}$) at recombination are unstable. The growth rate increases with wavenumber, until it saturates on very large wavenumbers: $k \approx 10^5 \ {\rm Mpc}^{-1}$, or physical wavelength $\lambda_{\rm phys} \approx 0.06 \ {\rm pc}$, or $10^4 \ {\rm AU}$. The modes at the saturation scale grow by a factor of a few hundred. Since there is a large number of small-scale modes, it is worth considering mechanisms that can cut off the growth on these scales. 

Photons in the continuum and \lya\ line interact strongly with matter during this epoch. We have briefly considered the aspects of this interaction relevant to background recombination in Section \ref{sec:background}. Continuum photons produced in direct recombinations to the ground state are completely unimportant for the background at the level of accuracy of Section \ref{sec:background}.  Their interaction cross section with neutral hydrogen atoms is so large that they are promptly reabsorbed. However, we should keep track of them in the in-homogenous case, since they can stream from one part of the fluctuation to another.

\begin{figure}[t]
  \includegraphics[width=\columnwidth]{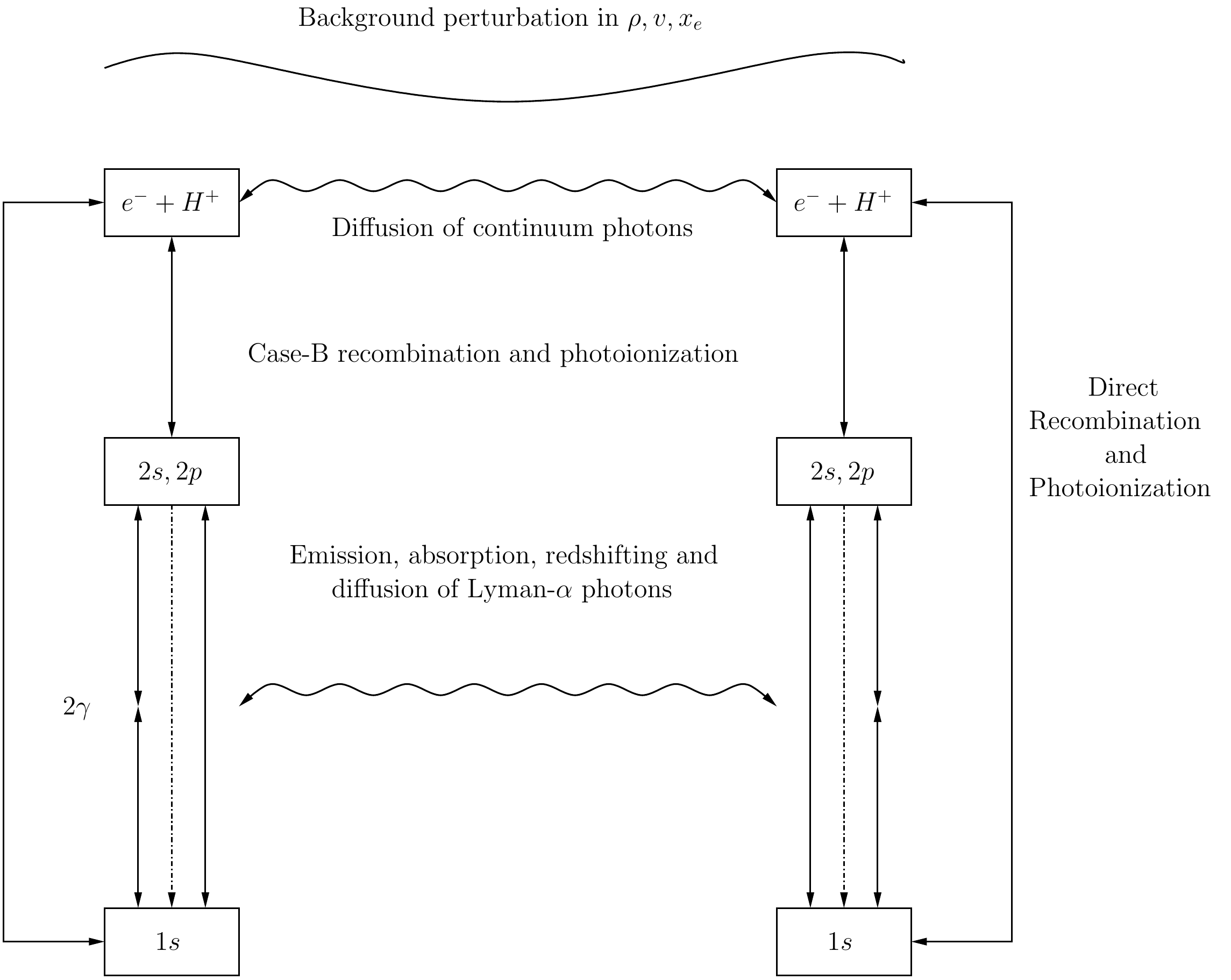}
  \caption{Schematic diagram showing the effect of continuum and \lya\ photon transport on the evolution of the ionization fraction fluctuation associated with small-scale fluctuations.}
  \label{fig:schem}
\end{figure}

Figure \ref{fig:schem} is a schematic diagram of the radiative transport processes relevant to perturbed recombination. Before we study the various processes in detail in subsequent sections, we clarify a few general points. 

Under the assumptions of the three level model of the hydrogen atom, we only need to consider a single spectral line (\lya). This greatly simplifies our analysis. The \lya\ photons can be decoupled from the continuum due to their wide separation in frequency. In the rest of this paper, we neglect the homogenous population of the first excited state, $x_2$ (except in equations which compute transitions from the $n=2$ level), and assume $x_{\rm e} + x_{1s} \approx 1$. As discussed in Section \ref{sec:background}, it is completely negligible compared to the other populations. 

A first step towards judging the relative importance of various arms of Fig.~\ref{fig:schem} is to look at the mean free paths (MFPs) of the photons at this redshift. If we use numbers for \lya\ photons at the line center, the comoving wavenumbers corresponding to the MFPs are
\begin{equation}
  k_{\rm cont} = n_{1s} \sigma_{1s, {\rm cont}} a \approx 3.6 \times 10^6 ~{\rm Mpc}^{-1}
\end{equation}
and
\begin{equation}
  k_{\lya} = n_{1s} \sigma_{1s,\lya} a \approx \frac{H \tau_{\rm S} a}{c \Delta_{\rm H}} \approx 10^{11} ~{\rm Mpc}^{-1} \mbox{.} \label{eq:naivediffusion}
\end{equation}
Here $\sigma_{1s, {\rm cont}} \approx 6.3 \times 10^{-18} ~{\rm cm}^2$ is the photo-ionization cross section for a ground state hydrogen atom at the threshold frequency, while $\tau_{\rm S} \approx 5.6 \times 10^8 $ and $\Delta_{\rm H} \approx 2.3 \times 10^{-5}$ are the Sobolev optical depth and the dimensionless Doppler width of the \lya\ line at the redshift of recombination. 

The MFP for continuum photons is very close to the saturation scale in Fig.~\ref{fig:growthsaha}. Moreover, as we show in Appendix \ref{subsec:diffusionscale}, the length scale for the diffusion of \lya\ photons is much larger than this naive estimate. In fact, we will see in Section \ref{sec:Lya} that \lya\ transport is important for wavenumbers satisfying $k > 10^{3} \ {\rm Mpc}^{-1}$. We begin by studying the outer arm of Fig.~\ref{fig:schem} in the next section. 

\section{Radiative transfer in the continuum}
\label{sec:continuum}

We study the transport of continuum photons in two stages -- we first determine their perturbed phase space density, and then calculate its effect on the recombination rate. We approach the problem using the Fourier-space Boltzmann equation (as used in previous sections and in modern CMB codes \cite{Ma1995, Seljak1996, CAMB, CLASS}). We note that the similar problem of ultraviolet and X-ray radiative transfer in the literature on high-redshift 21 cm radiation is usually addressed by a Green's function approach, i.e. by summing the contributions from individual point sources either analytically or numerically \cite{BarkanaLoeb05, PF07, Ahn09, Holzbauer12}.

Let the phase space density (henceforth, the PSD) of continuum photons be $f(\nu, \bm x, \hat{\bm n}, t)$. It evolves via the Boltzmann equation
\begin{align}
  \label{eq:boltzmann}
  \frac{\partial f}{\partial t}  - \Bigl[ H + \frac{n_i n_j}{a} \frac{\partial v_i}{\partial x_j} \Bigr] \nu \frac{\partial f}{\partial \nu} + \frac{c}{a} \hat{\bm n} \cdot \bm \nabla f = \sum_{\rm process} \dot{f} \vert_{\rm process} \mbox{.}
\end{align}
The second and third terms on the left-hand side account for the redshift of photons, and their advection respectively. Both the background cosmological expansion and the peculiar matter velocity contribute to the redshift term.

We assume that the PSD is not a dynamical variable and drop the explicit time-dependence. This is valid both in the unperturbed and perturbed cases: in the former, because photons redshift through the frequency range much faster than a Hubble time; and in the latter, because the advection term dominates below the Jeans scale.

We neglect the redshift term in Eq.~\eqref{eq:boltzmann}. This is equivalent to neglecting the background rate of recombination through the continuum channel. We consider the contributions of the absorption and emission of continuum photons to the right hand side of Eq.~\eqref{eq:boltzmann}, and neglect the redistribution of photons within the frequency range due to resonant scattering  -- this is important within the Lyman lines.

Let $\sigma_{\rm a}(\nu)$, $\alpha_{1s}(\nu)$ and $\phi(\nu)$ denote the continuum photon absorption cross-section, the direct recombination coefficient, and the probability distribution for the emitted photons' frequency respectively. These quantities are functions of radiation (absorption) and matter (recombination) temperature. The integrated or total recombination coefficient to the ground state is defined by 
\begin{equation}
  \alpha_{1s} = \int_{\nu_c}^{\infty} \!\! d\nu ~ \alpha_{1s}(\nu)\phi(\nu) \mbox{.}
\end{equation}
The rates of absorption and emission of continuum photons are
\begin{align}
  \dot{f}(\nu, \bm x, \hat{\bm n)} \vert_{\rm abs} &= -c n_{1s} \sigma_{\rm a}(\nu) f(\nu, \bm x, \hat{\bm n}) \mbox{,} \\
  \dot{f}(\nu, \bm x, \hat{\bm n}) \vert_{\rm em} &= \frac{c^3}{2 \nu^2} n_{\rm e} n_{\rm p} \alpha_{1s}(\nu) \frac{\phi(\nu)}{4\pi} \mbox{,}
\end{align}
where we have used the fact that every direct recombination is accompanied by the emission of a continuum photon, and multiplied by a factor of $c^3 n_{\rm H}/(2 \nu^2)$ to convert the contributions per hydrogen atom to those for the PSD. Substitution in the Boltzmann equation yields
\begin{align}
  ~~ & \!\!\!
  \frac{1}{a} \hat{\bm n} \cdot \bm \nabla f(\nu, \bm x, \hat{\bm n}) \nonumber \\
  &= - n_{1s} \sigma_{\rm a}(\nu) f(\nu, \bm x, \hat{\bm n}) + \frac{c^2}{8 \pi \nu^2} n_{\rm e} n_{\rm p} \alpha_{1s}(\nu) \phi(\nu) \mbox{.} \label{eq:boltzmanncont}
\end{align}
In the homogenous case, with just the background parameters, this reduces to the balance between absorption and recombination contributions.
\begin{align}
  0 &= \frac{1}{a} \hat{\bm n} \cdot \bm \nabla f(\nu) \\
  &= - (1- x_{\rm e}) n_{\rm H} \sigma_{\rm a}(\nu) f(\nu) + \frac{c^2}{8\pi \nu^2} (x_{\rm e} n_{\rm H})^2 \alpha_{1s}(\nu) \phi(\nu) \mbox{.} \label{eq:unpboltzmann1}
\end{align}
In the presence of small-scale fluctuations, we linearize the Boltzmann equation and simplify using the unperturbed solution, Eq.~\eqref{eq:unpboltzmann1}.
\begin{align}
  \frac{1}{a} \hat{\bm n} \cdot \bm \nabla \delta f (\nu, \bm x, \hat{\bm n}) + (1 - x_{\rm e}) n_{\rm H} \sigma_{\rm a}(\nu) \delta f(\nu, \bm x, \hat{\bm n} ) = \nonumber \\
  \frac{c^2}{8 \pi \nu^2} n_{\rm H}^2 x_{\rm e} \alpha_{1s}(\nu) \phi(\nu) \Bigl[ x_{\rm e} \delta_{\rm m} + \frac{2 - x_{\rm e}}{1 - x_{\rm e}} \delta x_{\rm e} \Bigr] \mbox{.} \label{eq:continuumrad}
\end{align}
Let the total number flux of continuum photons in a direction be $N(\bm x, \hat{\bm n})$. In terms of the PSD, it is given by 
\begin{equation}
  N(\bm x, \hat{\bm n}) = \int_{\nu_{\rm c}}^\infty \!\! d\nu \frac{8 \pi \nu^2}{c^2} f(\nu, \bm x, \hat{\bm n}) \mbox{.}
\label{eq:numberflux}
\end{equation} 
The photo-ionization cross-section, $\sigma_{\rm a}(\nu)$, is discontinuous across the threshold frequency. It falls off with increasing frequency in a power-law fashion \cite{Berestetskii71}, while the PSD falls in an exponential manner in the UV part of the spectrum. Hence we neglect the frequency dependence of $\sigma_{\rm a}$ in all integrals. Using Eq.~\eqref{eq:continuumrad} and the definition \eqref{eq:numberflux}, we get the equation for the transport of the number flux
\begin{equation}
  \frac{1}{a} \hat{\bm n} \cdot \bm \nabla \delta N(\bm x, \hat{\bm n}) + A \delta N(\bm x, \hat{\bm n}) = n_{\rm H} \Bigl[ B_1 \delta_{\rm m} + B_2 \delta x_{\rm e} \Bigr] \mbox{,} \label{eq:radequation2}
\end{equation}
where the coefficients are
\begin{subequations}
  \label{eq:coefficients}
  \begin{align}
    A &= (1 - x_{\rm e}) n_{\rm H} \sigma_{\rm a}(\nu_{\rm c}) \mbox{,} \\
    B_1 &= n_{\rm H} x_{\rm e}^2 \alpha_{1s} \mbox{,} \\
    B_2 &= n_{\rm H} x_{\rm e} \alpha_{1s} \frac{2 - x_{\rm e}}{1 - x_{\rm e}} \mbox{.}
  \end{align}
\end{subequations}
Note that the coefficient $A$ is the inverse of the mean free path for continuum photons at the threshold for photo-ionization.

We assume a plane-wave dependence for the fluctuation, following which the solution to Eq.~\eqref{eq:radequation2} is
\begin{equation}
  \frac{ \delta N(\bm k, \hat{\bm n}) }{ n_{\rm H} } = \frac{B_1 \delta_{\rm m} + B_2 \delta x_{\rm e}}{A + i (\hat{\bm n} \cdot {\bm k}/a)} \mbox{.} \label{eq:Isolution}
\end{equation}
The photo-ionization from and recombinations to the ground state together cause the free electron fraction to evolve as
\begin{align}
  \dot{x}_{\rm e} \vert_{\rm cont} &= x_{1s} \int_{\nu_{\rm c}}^\infty \!\! d\nu \frac{8 \pi \nu^2}{c^2} \sigma_{\rm a}(\nu) f_0(\nu, \bm x) - n_{\rm H} x_{\rm e}^2 \alpha_{1s} \mbox{.} \label{eq:xedotcontinuum}
\end{align}
In the homogenous case, we approximate the small contribution to be zero, which gives us a relation between the absorption cross-section and the recombination coefficient. 

We can obtain this relation by considering the alternative scenario of local thermal equilibrium (LTE) between a population of ionized and $1s$ hydrogens, free electrons, and a blackbody distribution of photons above the threshold frequency. The free electron fraction then equals the Saha equilibrium value of Eq.~\eqref{eq:saha}. As earlier, we neglect power-law frequency dependence of pre-factors in the integrals over frequency and obtain the relation
\begin{align}
  \alpha_{1s}(T) &= 4 \frac{h \nu_c}{m_{\rm e} c^2} \frac{h \nu_{\rm c}}{( 2 \pi m_e k_{\rm B} T)^{1/2}} \sigma_{\rm a}(\nu_{\rm c}, T) \mbox{.}
\end{align}
In the inhomogenous case, we perturb Eq.~\eqref{eq:xedotcontinuum} and retain terms up to the first order.
\begin{align}
  \delta \dot{x}_{\rm e} \vert_{\rm cont} &= x_{1s} \int_{\nu_{\rm c}}^\infty \!\! d\nu \frac{8 \pi \nu^2}{c^2} \sigma_{\rm a}(\nu) \Bigl[ \frac{\delta x_{1s}}{x_{1s}} f_0(\nu) + \delta f_0(\nu, \bm x) \Bigr] \nonumber \\
  & ~~~- n_{\rm H} x_{\rm e}^2 \Bigl[ \delta_{\rm m} + 2 \frac{\delta x_{\rm e}}{x_{\rm e}} \Bigr] \alpha_{1s} \mbox{.} 
\end{align}
We use detailed balance in the homogenous case, and the definition of the total flux in Eq.~\eqref{eq:numberflux} to simplify this contribution to
\begin{align}
  \delta \dot{x}_{\rm e} \vert_{\rm cont} &= (1 - x_{\rm e}) \sigma_{\rm a}(\nu_{\rm c}) \delta N_0 (\bm x) \nonumber \\
  & ~~~- n_{\rm H} x_{\rm e} \Bigl[ x_{\rm e} \delta_{\rm m} + \frac{2 - x_{\rm e}}{1 - x_{\rm e}} \delta x_{\rm e} \Bigr] \alpha_{1s} \\
  &= -\frac{1}{4\pi a n_{\rm H}} \int d\hat{\bm n} ~ \hat{\bm n} \cdot \bm \nabla \delta N(\bm x, \hat{\bm n}) \mbox{.} \label{eq:pxedotcontinuum}
\end{align}
To get to the second line, we used equation \eqref{eq:radequation2} for the the flux.

We use the solution \eqref{eq:Isolution} and evaluate the angular integral to obtain the final equation for the effect of continuum photon transport on the ionization fraction for a plane-wave fluctuation.
\begin{align}
  \delta \dot{x}_{\rm e} \vert_{\rm cont} & = - \Bigl\{ \frac{1}{4\pi} \int d\hat{\bm n} ~ \frac{i \hat{\bm n} \cdot \bm k }{A a + i \hat{\bm n} \cdot {\bm k}} \Bigr\} \Bigl[ B_1 \delta_{\rm m} + B_2 \delta x_{\rm e} \Bigr] \nonumber \\
  & = - \Bigl\{ 1 - \frac{A a}{k} \arctan{ \Bigl( \frac{k}{A a} \Bigr) } \Bigr\} \Bigl[ B_1 \delta_{\rm m} + B_2 \delta x_{\rm e} \Bigr] \mbox{,} \label{eq:xedotcont}
\end{align}
where the coefficients $A, B_1$ and $B_2$ are given in Eq.~\eqref{eq:coefficients}. The MFP of continuum photons is $1/A$; as expected the continuum photons' contribution goes to zero when the wavelength becomes much larger than this.

\section{Radiative transfer in Lyman-$\alpha$}
\label{sec:Lya}

This section works out the radiative transfer of \lya\ photons in an inhomogenous universe. The subject and details of this calculation are self-contained, but impact the rest of the paper through the resulting perturbed recombination rates. This sections' results are applicable over a wide range of length scales; we show that they reduce to expected values in the large- and small-scale limits in Appendices \ref{sec:levywalk} and \ref{sec:analytic}.  

The PSD of \lya\ photons evolves via the Boltzmann equation of Eq.~\eqref{eq:boltzmann}. It is simplest to work in the matter's rest frame, since the source terms on the right-hand side take on simple forms. Absorption, emission and resonant scattering contribute to this source term; we describe each of these processes in detail below.

The scattering of photons off a hydrogen atom in the ground state is a two step process, involving an excitation to a virtual excited state through the absorption of the incident photon, and subsequent decay through the emission of the outgoing one. When the first photon is of very low frequency, this corresponds to classical Rayleigh scattering. When its frequency approaches the \lya\ frequency (henceforth $\nu_{\lya}$), the intermediate state is long lived and other processes which deplete it become important. 

In particular, the excitation of the $2p$ state to higher bound states and its photo-ionization compete with spontaneous emission. We count the former as true absorptions, and the latter as coherent scattering events. The net photon number is unaffected by coherent scattering, but the frequency of the outgoing photon is related to that of the incident one. 

The branching ratio for coherent scattering is set by the rate of spontaneous emission from the $2p$ state
\begin{equation}
  p_{\rm sc} = \frac{A_{\lya}}{\Gamma_{2p}} = 1 - p_{\rm ab} \mbox{,} \label{eq:psc}
\end{equation}  
where $\Gamma_{2p}$ is the width due to all processes, and $p_{\rm ab}$ is the complementary branching ratio for absorption via two-photon processes. Coherent scattering is the dominant process, and the scattering probability $p_{\rm sc}$ is close to unity.

A useful definition is the Sobolev optical depth of the \lya\ line. It is the net optical depth for the absorption of a photon over its path as it redshifts through the \lya\ line due to cosmological expansion.
\begin{align}
  \tau_{\rm S} & = \frac{3}{8\pi} n_{1s} \left(\frac{c}{ \nu_{\lya}} \right)^3 \frac{A_{\lya}}{H} \mbox{.} \label{eq:sobolev}
\end{align}
The line is optically thick at the redshift of recombination, i.e. $\tau_{\rm S} \approx 5.6 \times 10^8 \gg 1$.  We divide this optical depth into true absorption and scattering contributions as
\begin{equation}
  \tau_{{\rm sc}/{\rm ab}} = p_{{\rm sc}/{\rm ab}} \tau_{\rm S} \mbox{.}
\end{equation}
The rate of removal of \lya\ photons per unit volume of phase space due to coherent scattering is
\begin{equation}
    \dot{f}(\nu, \hat{\bm n}) \vert_{{\rm sc}-} = - H \nu \tau_{\rm sc} \phi(\nu) e^{[h(\nu-\nu_{\lya})/k_{\rm B} T]} f(\nu, \hat{\bm n}) \mbox{.} \label{eq:lyasc-}
\end{equation}
In the above expression, $\phi(\nu)$ is broadened from a delta function at the \lya\ frequency, $\nu_{\lya}$, due to the thermal motions of the absorbing atoms and the finite lifetime of the excited state. The resulting profile is a Voigt function, which is most easily expressed in terms of the deviation from the central frequency in Doppler widths \cite{Hummer1962}:
\begin{align}
  \phi (x,a) & = \frac{a}{\pi^{3/2}} \int_{-\infty}^\infty du \frac{e^{-u^2}}{a^2 + (x-u)^2} \mbox{,} \\
  x & = \frac{\nu - \nu_{\lya}}{\nu_{\lya} \Delta_{\rm H}}, \qquad \Delta_{\rm H} = \biggl( \frac{2k_{\rm B} T}{m_{\rm H} c^2} \biggr)^{1/2} \mbox{.}
\end{align}
The Voigt parameter, $a$, quantifies the relative strength of the radiative and Doppler broadening mechanisms, and is given by
\begin{equation}
  a = \frac{\Gamma_{2p}}{4\pi \nu_{\lya} \Delta_{\rm H}} \mbox{.}
\end{equation}
The outgoing photon follows a redistribution function, $p(\nu, \hat{\bm n} \vert \nu^\prime, \hat{\bm n}^\prime)$. This is defined as the probability of an outgoing photon $(\nu, \hat{\bm n})$ conditioned on the incoming photon $(\nu^\prime, \hat{\bm n}^\prime)$ \cite{Hummer1962}. It is normalized as
\begin{equation}
  \int d\nu \frac{d\hat{\bm n}}{4\pi} p(\nu, \hat{\bm n} \vert \nu^\prime, \hat{\bm n}^\prime) = 1 \mbox{.}
\end{equation}
The rate of injection per unit volume of phase space due to coherent scattering is 
\begin{align}
  \dot{f}(\nu, \hat{\bm n}) \vert_{{\rm sc}+} & = H \nu \tau_{\rm sc} \int d\nu^\prime \frac{d\hat{\bm n}^\prime}{4\pi} \phi(\nu^\prime) e^{[h(\nu^\prime - \nu_{\lya})/k_{\rm B} T]} \nonumber \\
  & ~~~\times p(\nu, \hat{\bm n} \vert \nu^\prime, \hat{\bm n}^\prime ) f(\nu^\prime, \hat{\bm n}^\prime) \mbox{.} \label{eq:lyasc+}
\end{align} 
True absorptions are two-photon transitions to higher states, through an intermediate `virtual' $2p$ state. Direct photo-ionization from the $2p$ state is formally included by letting the summation over the higher states run over the continuum states. However, the dominant transitions from $2p$ are to the $3s$ and $3d$ levels. To the first approximation, the resultant absorption probability is
\begin{align}
  p_{\rm ab} & \approx \frac{A_{3s-2p} + 5 A_{3d-2p}}{3 A_{\lya}} e^{- (5 h \nu_{\lya}/ 27 k_{\rm B} T) } \nonumber \\
  & \approx 10^{-4} \ {\rm at} \ z_0 = 1100 \mbox{.} \label{eq:pab}
\end{align}
In the first line, we have neglected the absorption contribution in the denominator, and assumed that the PSD for the second photon of lower energy is that of a blackbody at the radiation temperature. The rate of removal of photons due to true absorption is
\begin{equation}
  \dot{f}(\nu, \hat{\bm n}) \vert_{\rm ab} = - H \nu \tau_{\rm ab} \phi(\nu) e^{[h (\nu-\nu_{\lya})/k_{\rm B} T]} f(\nu, \hat{\bm n}) \mbox{.} \label{eq:lyaab}
\end{equation}
In a similar manner, true emission of \lya\ photons is a two-photon process, in which the first photon is emitted in a transition from one of the higher levels (as earlier, largely from $3s$ and $3d$) to a `virtual' $2p$ level, and the second one during a subsequent decay to the ground state. We neglect the stimulated component of both transitions since the PSDs involved are much smaller than unity. The rate of injection due to true emission is
\begin{equation}
  \dot{f}(\nu, \hat{\bm n}) \vert_{\rm em} = \frac{c^3 n_{\rm H}}{8 \pi \nu^2} p_{\rm sc} \sum_{i \ne 1s} x_i A_{i-2p} \phi(\nu)  \label{eq:emission} \mbox{.}
\end{equation}

\begin{figure}[t]
  \includegraphics[width=\columnwidth]{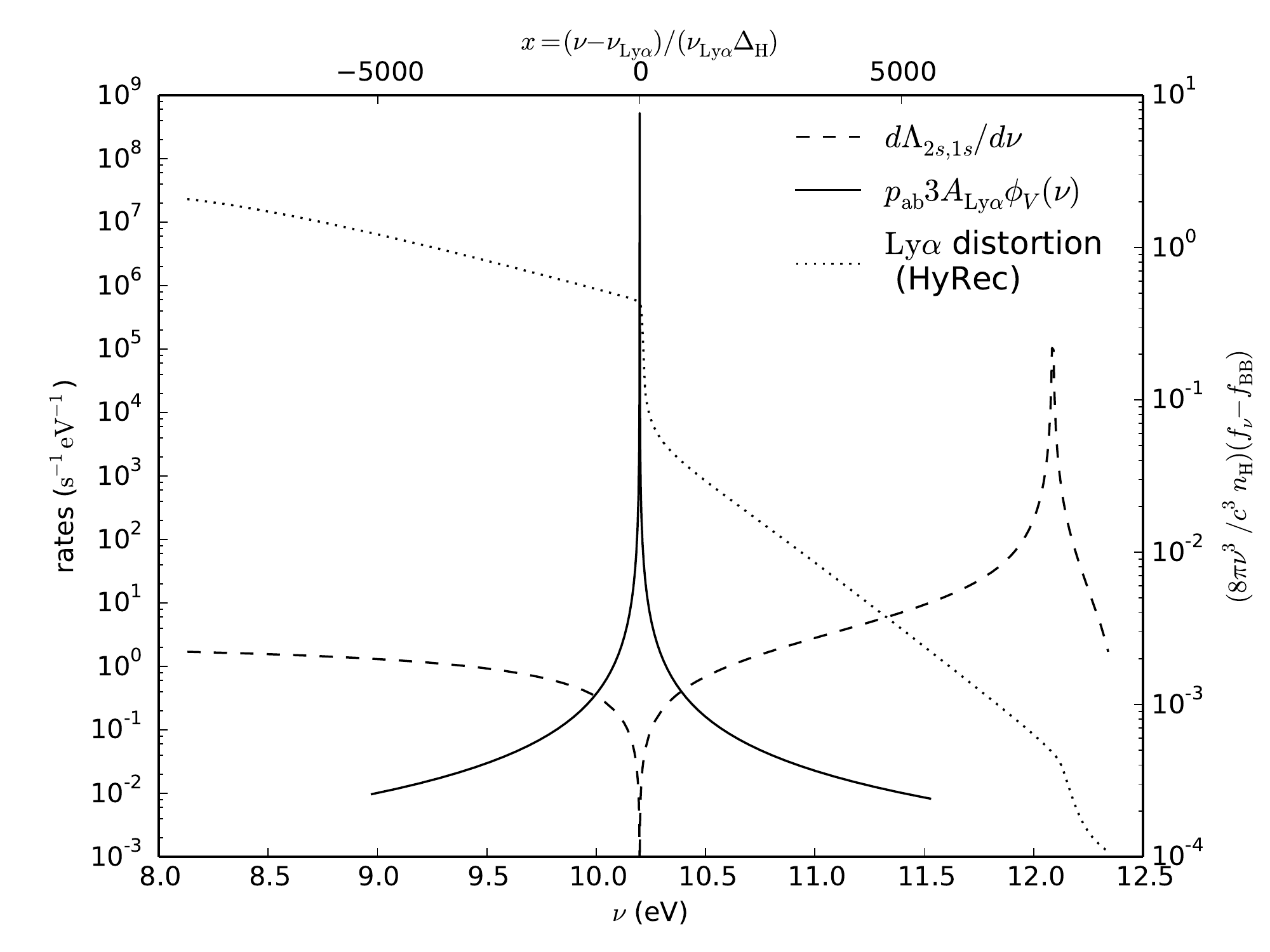}
  \caption{Rates of radiative processes: The solid and dashed lines show the rate coefficients per unit frequency for two-photon absorption via an intermediate $2p$ level, and two-photon absorption/Raman scattering to the $2s$ level, respectively. The lower and upper axes show the frequency in physical units, and Doppler widths from line center respectively. Also shown on the same plot is the spectral distortion, as calculated by HyRec \cite{HyRec}. The dotted line shows the number of excess photons over a blackbody per hydrogen atom per logarithmic frequency interval. The plots are generated at redshift $z_0 = 1100$.}
  \label{fig:rates}
\end{figure}

In principle, two-photon transitions to and from the $2s$ state can also inject or remove photons within the \lya\ line. Depending on the frequency of the more energetic photon involved, these are Raman scattering or two-photon transitions between $2s$ and the ground state. However, these transitions are much slower than those involving the $2p$ state; in particular, their cross-section goes to zero at the central frequency, since there is no phase space available for the second photon (see Fig.~\ref{fig:rates}). This statement is no longer true if we include stimulated emission, but the full transition rates are still much smaller than the ones to $2p$ within the \lya\ line \cite{TwoPhoton}. Thus, the majority of photons produced in this manner are on the far red side of the line. We can safely neglect this channel while calculating the spectral distortion within a few hundred Doppler widths of the \lya\ line center.

Fig.~\ref{fig:rates} shows the rates of the radiative processes described above which add or remove photons from the frequency range of interest.

\subsection{Solution of the Boltzmann equation}
\label{subsec:solutionassumptions}

We solve the Boltzmann equation under a number of simplifying assumptions.
\begin{enumerate}
 \item The $2s$--$2p$ transition rate is high enough so that all their sublevels are equally occupied. Consequently we neglect the fast transitions between these sublevels.
 \item The line profile, $\phi(\nu)$, dominates the frequency dependence of the absorption and emission terms. Thus we replace all factors of $\nu$ multiplying the profile with the central frequency, $\nu_{\lya}$.
 \item The rates of radiative processes are large compared to the Hubble rate, so the PSD and excited level populations are effectively in steady state. This is valid within the line profile due to the high scattering rate.
 \item The absorption and emission profiles are identical. Under this approximation, factors of $\exp{[h(\nu-\nu_{\lya})/k_{\rm B} T]}$ are approximately equal to unity. This is valid if we restrict ourselves to frequencies which satisfy
   \begin{align}
     \vert \nu - \nu_{\lya} \vert & \ll \nu_{\lya} \Delta_{\rm H} X \mbox{,} \\
     X & = \frac{k_{\rm B} T}{h \nu_{\lya} \Delta_{\rm H}} \approx 1080, ~ {\rm at} ~ z_0 = 1100 \nonumber \mbox{.}
   \end{align}
   This is satisfied within the frequency range of interest, since the wings are optically thick to true absorption only up to $\sim 20$ Doppler widths at this redshift \cite{AHGH2010}.
 \item On the far blue side of the line, we take the PSD to equal that of a blackbody at the radiation temperature. 
 \item The redistribution function, $p(\nu, \hat{\bm n} \vert \nu^\prime, \hat{\bm n}^\prime)$, is isotropic. We condense it to the the notation $p(\nu \vert \nu^\prime)$.
\end{enumerate}
We use the steady state approximation to balance the rate of processes which populate the $2p$ level -- downward transitions from higher levels and upward transitions from the $1s$ level -- with its net rate of depletion. 
\begin{align}
  0 & = \dot{x}_{2p} \nonumber \\
  & = \sum_{i \ne 1s} x_i A_{i-2p} + 3 x_{1s} A_{\lya} \overline{f} - \Gamma_{2p} x_{2p} \mbox{.} \label{eq:2psteady}
\end{align}
where $\overline{f}$ is the average of the phase-space density over the line profile, $\overline{f} = \int d\nu \ \phi(\nu) f(\nu)$. We use this along with the definition of the scattering probability in Eq.~\eqref{eq:psc} to rewrite the emission term of Eq.~\eqref{eq:emission} as
\begin{equation}
  \dot{f}(\nu, \hat{\bm n}) \vert_{\rm em} = H \nu \tau_{\rm S} \phi(\nu) \Bigl[ f_{\rm eq} - p_{\rm sc} \overline{f} \Bigr] \mbox{,} \label{eq:lyaem}
\end{equation}
where we have introduced the equilibrium PSD, $f_{\rm eq}$, which is defined as
\begin{align}
  f_{\rm eq} = \frac{x_{2p}}{3 x_{1s}} = \frac{x_2}{4 x_{1s}} \mbox{.} \label{eq:feq}
\end{align}

\subsubsection{Homogenous case}
\label{subsubsec:homogenous}

If the background ionization state and density are homogenous, the PSD is independent of direction and position. Under the assumptions listed above, the Boltzmann equation of Eq.~\eqref{eq:boltzmann} reduces to 
\begin{align}
  \frac{\partial f (\nu)}{\partial \nu} & = \tau_{\rm sc} \Bigl[ \phi(\nu) f(\nu) - \int d\nu^\prime \phi(\nu^\prime) p(\nu^\prime, \nu ) f(\nu^\prime) \Bigr] \nonumber \\
  & ~~~+ \tau_{\rm S} \phi(\nu) \Bigl[ p_{\rm ab} f(\nu) - f_{\rm eq} + p_{\rm sc} \overline{f} \Bigr] \mbox{.} \label{eq:lyrad0} 
\end{align}
This is easily solved if the redistribution due to coherent scattering is unimportant, i.e., $p_{\rm sc} \approx 0$, or independent of the incoming frequency, i.e., $p(\nu^\prime, \nu) = \phi(\nu)$. The PSD is then given by the Sobolev solution. Complete redistribution is a good approximation within the Doppler core (up to $\sim 40$ Doppler widths away from $\nu_{\lya}$ at $z = 1100$ \cite{AHGH2010}). 

However, redistribution due to coherent scattering is nontrivial in the wings, since the average change in frequency between the incident and outgoing photons is only a few Doppler widths. We implement the resulting diffusion in frequency using a second-order differential operator. This is commonly known as the Fokker-Planck approximation \cite{Unno1955, Rybicki1994, AHGH2010}. It is well suited for describing the partial redistribution in the wings. Due to the high scattering rates near the line center, the PSD sets itself to the equilibrium value, and the particular prescription used becomes unimportant, as long as it yields a small result. Under this approximation, the rates of injection and removal due to scattering are
\begin{align}
  \dot{f}(\nu) \vert_{\rm sc} & = - H \nu \tau_{\rm sc} \Bigl[ \phi(\nu) f(\nu) - \int d\nu^\prime \phi(\nu^\prime) p(\nu^\prime, \nu ) f(\nu^\prime) \Bigr] \nonumber \\
  & = H \nu \tau_{\rm sc} \frac{\nu_{\lya}^2 \Delta_{\rm H}^2}{2} \frac{\partial}{\partial \nu} \Bigl[ \phi(\nu) \frac{\partial f}{\partial \nu} \Bigr] \mbox{,}
\end{align}
The operator above does not account for the effect of atomic recoil; this is consistent with the approximation of equal absorption and emission profiles (assumption 4). Using this in Eq.~\eqref{eq:lyrad0}, we get a second--order ordinary differential equation (ODE) for the phase-space density
\begin{align}
  \frac{\partial f (\nu) }{\partial \nu} & = - \tau_{\rm sc} \frac{\nu_{\lya}^2 \Delta_{\rm H}^2}{2} \frac{\partial}{\partial \nu} \Bigl[ \phi(\nu) \frac{\partial f}{\partial \nu} \Bigr] \nonumber \\
  & ~~~+ \tau_{\rm S} \phi(\nu) \Bigl[ p_{\rm ab} f(\nu) - f_{\rm eq} + p_{\rm sc} \overline{f} \Bigr] \mbox{.} \label{eq:lyrad0fp}
\end{align}
We numerically solve this differential equation in a frequency range extending out to $1000$ Doppler widths on either side of $\nu_{\lya}$, with $50$ bins per Doppler width. We set the PSD to a blackbody on the far blue side, and use a Neumann boundary condition on the far red side, where we set the derivative to zero. The latter is designed to kill an unphysical solution where the PSD grows catastrophically as we approach the red side of the line.

Technically, this region is larger than the domain of validity for some of our approximations, but we formally extend the equation out to this region in order to reduce boundary effects. We evaluate the Voigt profile using Gubner's series in the core, and a fourth order asymptotic expansion in the wings \cite{Gubner94}. 

In order to evaluate the equilibrium PSD, $f_{\rm eq}$, we need the occupancies of the ground ($1s$) and excited ($2p$) states. The rates of their depletion and population depend on the PSD itself, so to be completely self-consistent, we need to solve for the level populations together with the PSD. Instead, we use the three level model of recombination of Section \ref{sec:background}, which assumes the Sobolev solution. The error introduced by doing so is small, because the most significant effect of the redistribution is to broaden the jump in the PSD, rather than change its amplitude.

\begin{figure}[t]
  \includegraphics[width=\columnwidth]{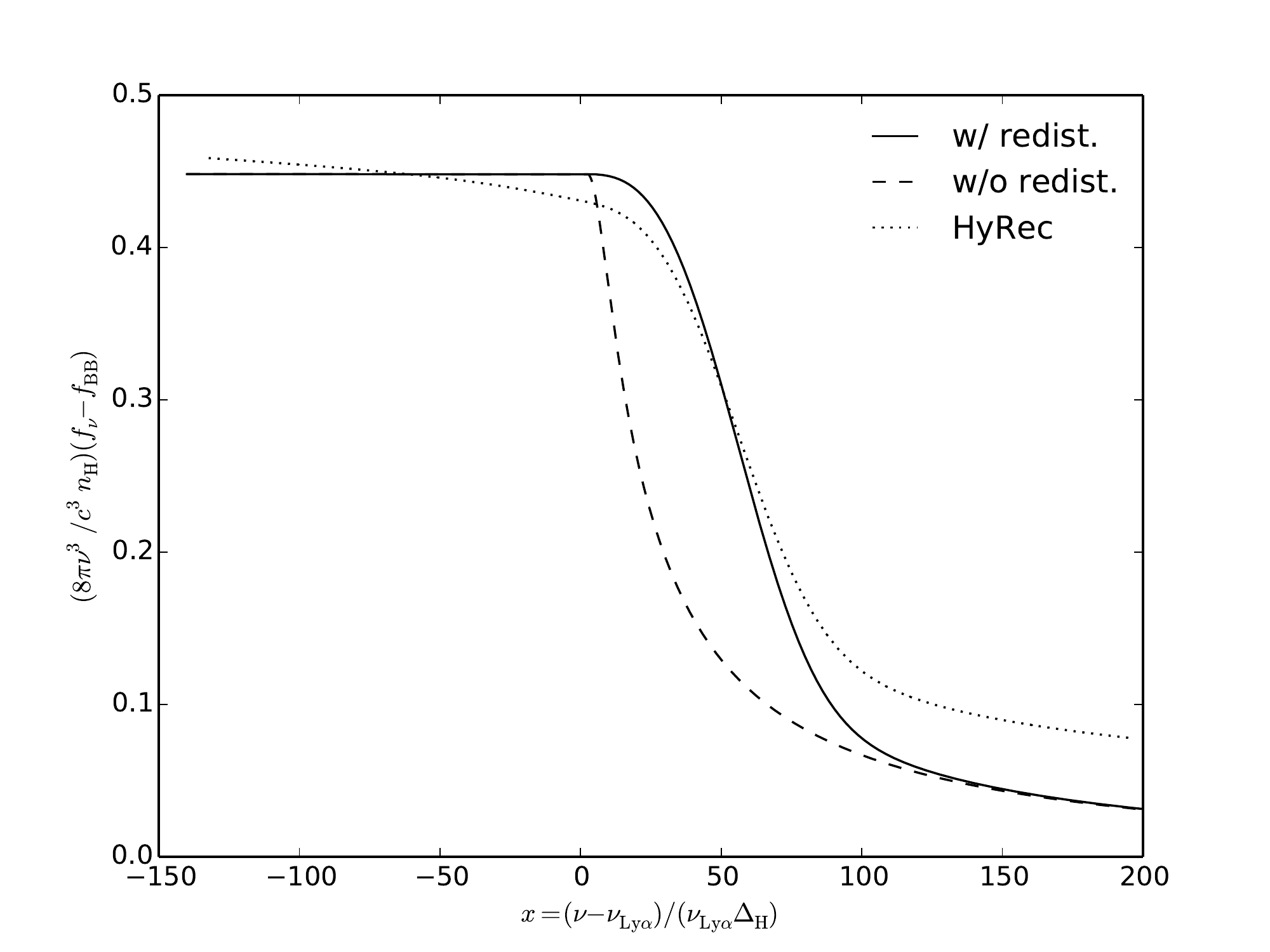}
  \caption{ \lya\ spectral distortion: This figure plots the number of excess photons over a blackbody distribution per hydrogen atom per logarithmic frequency interval, against the frequency offset from line center measured in Doppler widths, at redshift $z_0 = 1100$. The solid line is the solution of Eq.~\eqref{eq:lyrad0fp}, which incorporates redistribution due to coherent scattering, while the dashed one is the Sobolev solution, which does not. Also shown for reference is the result of the full calculation of HyRec \cite{HyRec}. }
  \label{fig:homogenous}
\end{figure}

Figure \ref{fig:homogenous} shows the resulting spectral distortion, which is defined via the PSD as the number of excess photons over a blackbody distribution per hydrogen atom per logarithmic frequency interval. Also shown are the true distortion (as calculated by the publicly available HyRec code \cite{HyRec}), and the Sobolev approximation to it, which neglects redistribution due to coherent scattering. HyRec's treatment of recombination and radiative processes is significantly more sophisticated than ours -- it does not assume a steady state or equal emission and absorption profiles, follows the population of the higher levels, and accounts for two-photon and Raman transitions which are nonresonant with the \lya\ transition.

The rate of recombination through the \lya\ channel is the difference between the downward and upward transition rates 
\begin{align}
  \dot{x}_{1s} \vert_{\lya} & = 3 A_{\lya} x_{1s} \Bigl[ f_{\rm eq} - \overline{f}_{0 0} \Bigr] \mbox{.} \label{eq:lyarec}
\end{align}
We get an expression for the average monopole, $\overline{f}_{0 0}$, and hence the recombination rate through the Ly$\alpha$ channel by integrating Eq.~\eqref{eq:lyrad0} over frequency, and using the normalization of the redistribution probability. 
\begin{equation}
  \Delta f = \tau_{\rm S} [ \overline{f} - f_{\rm eq} ] \mbox{,}
\end{equation}
where the notation $\Delta X$ respresents the jump in a quantity $X$ across the line, $\Delta X = X(\nu_+) - X(\nu_-)$. Using this in Eq.~\eqref{eq:lyarec}, we recover the background recombination rate in the Sobolev approximation with large optical depth
\begin{equation}
  \dot{x}_{1s} \vert_{\lya} = - \frac{3 A_{\lya} x_{1s}}{\tau_{\rm S}} \Delta f \mbox{.} \label{eq:xedotsobolev}
\end{equation}
Typically the PSD on the red side, $f_{\nu_-}$, sets itself to the equilibrium value, $f_{\rm eq}$, due to the high optical depth. On the far blue side, we take $f_{\nu_{+}}$ to equal the blackbody value to maintain consistency with assumption 5 and the numerical solution.

A significant fraction of atoms reach the ground state via two-photon decays from the $2s$ level. From Fig.~\ref{fig:rates}, we see that the more energetic of the emitted photons is largely on the far red side of the \lya\ line. The effect of absorption of the background spectral distortion in this region is largely canceled by that of the stimulated emission of the low energy photon \cite{TwoPhoton}. Thus, we compute the two-photon decay rate using the blackbody PSD.
\begin{subequations}
  \label{eq:twophoton}
  \begin{align}
    \dot{x}_{1s} \vert_{2s} & = \Lambda_{2s,1s} x_{1s} \Bigl[ f_{\rm eq} - e^{-\{ h \nu_{\lya}/k_{\rm B} T \}} \Bigr] \mbox{,} \\
    \Lambda_{2s,1s} & = \int_{\nu_{\lya}/2}^{\nu_{\lya}} d\nu \frac{d\Lambda_{2s}}{d\nu} = 8.22 ~{\rm s}^{-1} \mbox{.}
  \end{align}
\end{subequations}
We neglect Raman scattering events involving photons above $\nu_{\lya}$. Their main impact on recombination is `nonlocal' in time; they inject photons on the far blue side of \lya\, which redshift into the line at a later time due to cosmological expansion and get absorbed \cite{TwoPhoton}.

Equations \eqref{eq:xedotsobolev} and \eqref{eq:twophoton} together give the net rate of recombination to the ground state. The result depends on the equilibrium PSD, $f_{\rm eq}$, which in turn depends on the $n=2$ level's population. We use the steady state assumption and balance its overall rates of population and depopulation. 

One way of implementing this would be to follow the populations of all the levels which connect to it, in the manner of Eq.~\eqref{eq:2psteady}. Instead, we choose to work in the three level approximation of Section \ref{sec:background}, which collects all the higher levels into a single block and assumes equal population for all the sublevels. The rates of case B recombination and photo-ionization add up to give the rate of the upper arms, which connect the fully ionized state with the $n=2$ state.
\begin{equation}
  \dot{x}_2 \vert_{\rm rec/ion} = n_{\rm H} x_{\rm e}^2 \alpha_{\rm B} - x_2 \beta_{\rm B} \mbox{.} \label{eq:upperarm}
\end{equation}
If we equate this expression to the sum of Eqs.~\eqref{eq:xedotsobolev} and \eqref{eq:twophoton}, we recover Eq.~\eqref{eq:peebles} after some algebra. The explicit expressions for Peebles' $C$ factor and the $n=2$ population are
\begin{align}
  C & = \frac{ 3 A_{\lya}/\tau_{\rm S} + \Lambda_{2s,1s} }{ 3 A_{\lya}/\tau_{\rm S} + \Lambda_{2s,1s} + 4 \beta_{\rm B} } \mbox{,} \label{eq:peeblescfactor} \\
  x_2 & = 4 \frac{n_{\rm H} x_{\rm e}^2 \alpha_{\rm B} + (3 A_{\lya}/\tau_{\rm S} + \Lambda_{2s,1s})x_{1s} e^{-E_{21}/(k_{\rm B} T)}}{ 3 A_{\lya}/\tau_{\rm S} + \Lambda_{2s,1s} + 4 \beta_{\rm B}} \mbox{.} \label{eq:x2}
\end{align}

\subsubsection{Inhomogenous case}
\label{subsec:inhomogenous}

The situation of interest in this paper involves spatially varying hydrogen number density, ionization fraction and matter velocity. The resulting phase-space density in \lya\ is both inhomogenous, i.e., varies with position $\bm x$, and anisotropic, i.e., varies with direction $\hat{\bm n}$. We assume that these variations take the form of small fluctuations over a homogenous background, so that we can expand their spatial dependence into plane waves which evolve independently of each other. They obey the Boltzmann equation \eqref{eq:boltzmann}, whose linearized form is
\begin{align}
  ~~ & \!\!\! 
  \frac{\partial \delta f}{\partial \nu} - \frac{i c k}{H \nu a} (\hat{\bm k} \cdot \hat{\bm n}) \delta f  - \frac{\delta \tau_{\rm S}}{\tau_{\rm S}} \frac{\partial f}{\partial \nu} \nonumber \\
  & = \tau_{\rm S} \phi(\nu) \Bigl[ p_{\rm ab} \delta f(\nu, \hat{\bm n}) - \delta f_{\rm eq} + p_{\rm sc} \delta \overline{f}_{0 0} \Bigr] \nonumber \\
  & ~~~+ \tau_{\rm sc} \Bigl[ \phi(\nu) \delta f(\nu, \hat{\bm n}) - \int d\nu^\prime \frac{d\hat{\bm n}^\prime}{4\pi} \phi(\nu^\prime) p(\nu \vert \nu^\prime ) \delta f(\nu^\prime, \hat{\bm n}^\prime) \Bigr] \mbox{.} \label{eq:linboltzmann}
\end{align}
Here the perturbed source terms on the right hand side include the effects of absorption [Eq.~\eqref{eq:lyaab}], emission [Eq.~\eqref{eq:lyaem}], and scattering [Eqs.~\eqref{eq:lyasc-} and \eqref{eq:lyasc+}], after applying the assumptions listed at the beginning of Section \ref{subsec:solutionassumptions}.

\begin{figure*}
\centering
\subfloat[][Source: $\delta_{\rm m} + \delta x_{1s}/x_{1s}$]{
    \includegraphics[width=0.5\textwidth]{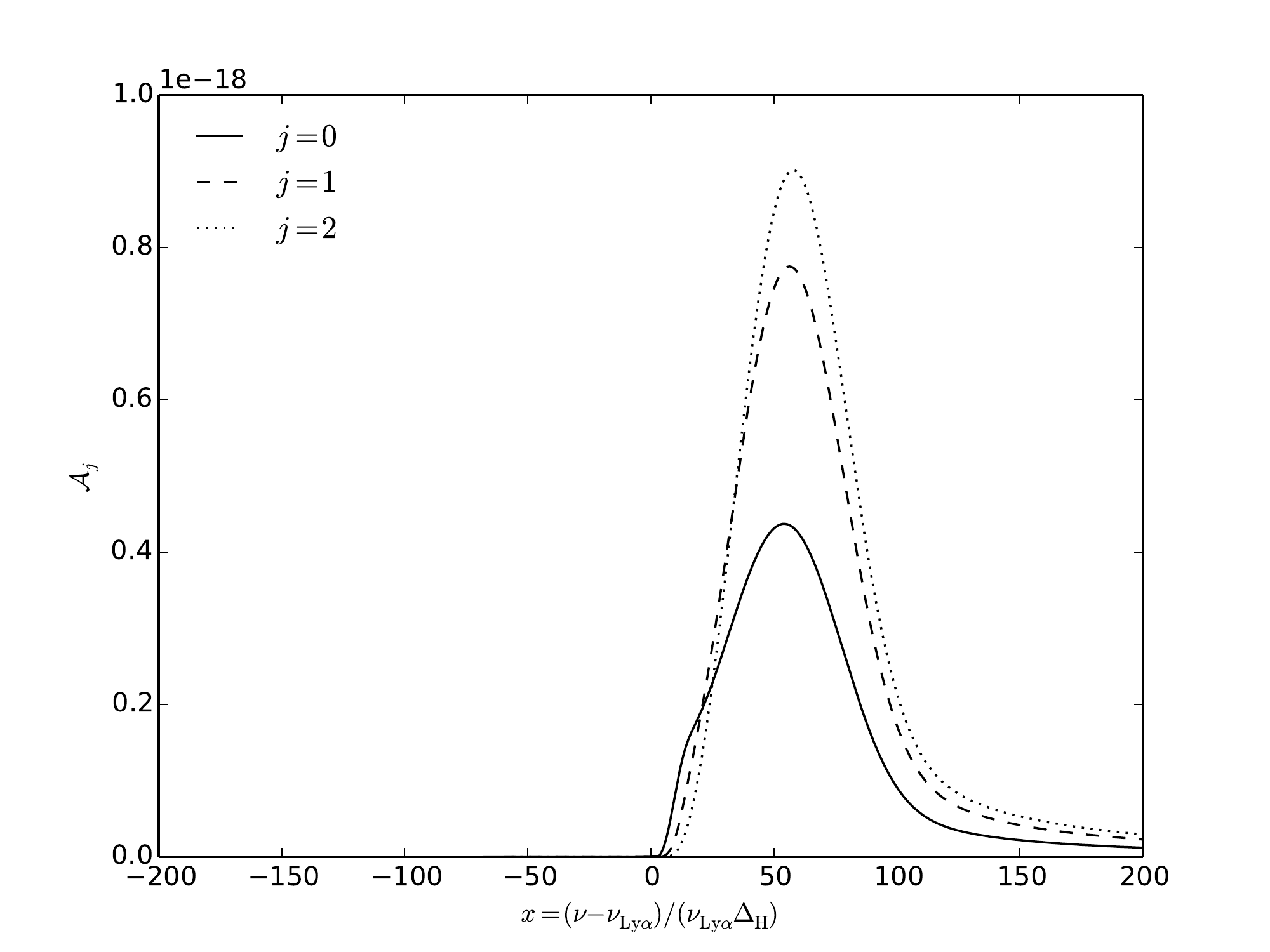}
    \label{fig:sourcenhx1s}
    }
    \subfloat[][Source: $\Theta/a H$]{
    \includegraphics[width=0.5\textwidth]{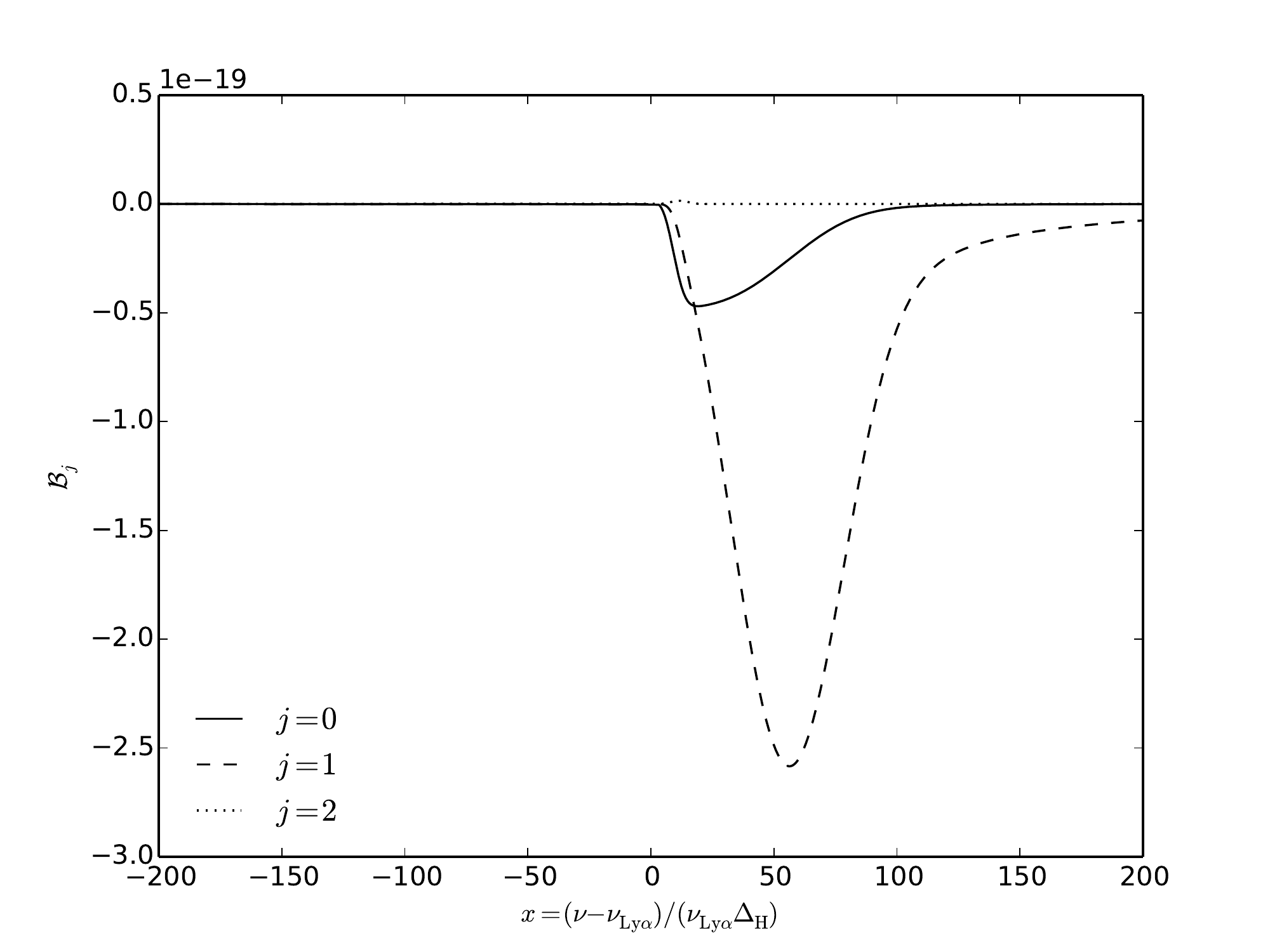}
    \label{fig:sourceTheta}
    }
    \caption{Basis solutions for the inhomogenous Boltzmann equation: (a) and (b) show the solutions $\mathcal{A}_j$ and $\mathcal{B}_j$ defined in Eq.~\eqref{eq:basissolns}. The source terms modulate the optical depth through the density and ground state population, and velocity gradient. This figure is generated for $k=10^{5} ~ {\rm Mpc}^{-1}$ at redshift $z_0=1100$.}
  \label{fig:sourceAB}
\end{figure*}

The fluctuation in the optical depth is
\begin{equation}
  \frac{ \delta \tau_{\rm S} }{ \tau_{\rm S} } = \delta_{\rm m} + \frac{ \delta x_{1s} }{ x_{1s} } - \frac{ \Theta }{ a H } ( \hat{\bm k} \cdot \hat{\bm n} )^2 \mbox{.}
\end{equation}
We decompose the angular dependence of quantities into their spherical harmonic components. (This -- or some more sophisticated variant -- is the standard approach for Boltzmann solvers that predict CMB anisotropies \cite{Ma1995, Seljak1996, Hu1997, CAMB, CLASS}.) It is convenient to orient the z-axis, $\hat{\bm z}$, along the wave vector, $\bm k$. Due to azimuthal symmetry about this axis, quantities depend on direction only through $\mu = \hat{\bm k} \cdot \hat{\bm n}$, and the spherical harmonics reduce to the appropriate Legendre polynomials. The explicit forms of the decomposition and its inverse for the PSD are \cite{Ma1995, Hu1997}
\begin{subequations}
\label{eq:moments}
\begin{align}
  \delta f(\nu, k, \mu) & = \sum_{j} (-i)^j \delta f_j(\nu, k) P_{j}(\mu) \mbox{,} \\
  \delta f_j(\nu, k) & = \frac{2j+1}{2} \int d \mu \ i^j P_{j}(\mu) \delta f(\nu, k, \mu) \mbox{.}
\end{align}
\end{subequations}
We substitute the expansion \eqref{eq:moments} into Eq.~\eqref{eq:linboltzmann} to get the Boltzmann equations for the moments. The equation for the zeroeth moment is
\begin{align}
  \frac{\partial \delta f_{0}}{\partial \nu} & = \frac{\delta \tau_{\rm S, 0}}{\tau_{\rm S}} \frac{\partial f}{\partial \nu} - \tau_{\rm S} \phi(\nu) \Bigl[ \delta f_{\rm eq}  - p_{\rm sc} \delta \overline{f}_{0} \Bigr]  \nonumber \\
  & ~~~+ \frac{c k}{3 H \nu a} \delta f_{1} + p_{\rm ab} \tau_{\rm S} \phi(\nu) \delta f_{0} + \tau_{\rm sc} \Bigl\{ \phi(\nu) \delta f_{0} \nonumber \\
  & ~~~- \int d\nu^\prime \phi(\nu^\prime) p(\nu | \nu^\prime) \delta f_{0}(\nu^\prime) \Bigr\} \mbox{.}
\end{align}
The term within curly braces is the scattering contribution, which redistributes photons within the line. We replace it with a second-order differential operator under the Fokker-Planck approximation, in the same manner as in the homogenous case.
\begin{align}
  \frac{\partial \delta f_{0}}{\partial \nu} & = \frac{\delta \tau_{\rm S, 0}}{\tau_{\rm S}} \frac{\partial f}{\partial \nu} - \tau_{\rm S} \phi(\nu) \Bigl[ \delta f_{\rm eq}  - p_{\rm sc} \delta \overline{f}_{0} \Bigr] + \frac{c k}{3 H \nu a} \delta f_{1} \nonumber \\
  & ~~~+ p_{\rm ab} \tau_{\rm S} \phi(\nu) \delta f_{0} - \tau_{\rm sc} \frac{\nu_{\lya}^2 \Delta_{\rm H}^2}{2} \frac{\partial}{\partial \nu} \Bigl[ \phi(\nu) \frac{\partial \delta f_0}{\partial \nu} \Bigr] \mbox{.} \label{eq:monopole}
\end{align}
The Boltzmann equations for the higher moments, with $j \ge 1$, are of the form
\begin{align}
  \frac{\partial \delta f_{j}}{\partial \nu} & = \frac{c k}{H \nu a} \Bigl[ - \frac{j}{2j-1} \delta f_{j-1} + \frac{j+1}{2j+3} \delta f_{j+1} \Bigr] \nonumber \\
  & ~~~+ \tau_{\rm S} \phi(\nu) \delta f_{j} + \frac{\delta \tau_{\rm S, 2}}{\tau_{\rm S}} \frac{\partial f}{\partial \nu} \delta_{j,2} \mbox{,} \label{eq:jge1boltzmann}
\end{align}
where the $\delta_{j,2}$ in the final term on the RHS equals unity if $j=2$ and zero otherwise.

Equations \eqref{eq:monopole}--\eqref{eq:jge1boltzmann} form a hierarchy for the moments of the PSD, $\delta f_{j 0}$ \cite{Ma1995, Seljak1996}. Absorption, emission and redshifting of \lya\ photons contribute to the evolution of each moment, while redistribution due to coherent scattering only contributes to the zeroeth moment. The latter is a direct consequence of the assumption of the isotropy of the redistribution function, $p(\nu, \hat{\bm n} \vert \nu^\prime, \hat{\bm n}^\prime)$ (assumption 6). In addition to this, free-streaming couples moments whose angular indices differ by unity \cite{Hu1997}.

We obtain the complete solution by adding the ones for each of the source terms as follows:
\begin{align}
  \delta f_{j}(\nu) & = \biggl( \delta_{\rm m} + \frac{ \delta x_{1s} }{ x_{1s} } \biggr) \mathcal{A}_{j} (\nu) + \frac{\Theta}{a H} \mathcal{B}_{j} (\nu) \nonumber \\
  & ~~~+ \Bigl( \delta f_{\rm eq} - p_{\rm sc} \delta \overline{f}_{0} \Bigr) \mathcal{C}_{j} (\nu) \mbox{,} \label{eq:basissolns}
\end{align}
where $\mathcal{A}_j, \mathcal{B}_j$ and $\mathcal{C}_j$ are dimensionless solutions sourced by combinations of the first and second terms on the RHS of Eq.~\eqref{eq:monopole}, and the last term on the RHS of \eqref{eq:jge1boltzmann}. The notation for $\mathcal{C}_j$ is used only in this section, and is not to be confused with Peebles' $C$ factor. 

We numerically solve the Boltzmann hierarchy of Eq.~\eqref{eq:monopole} and \eqref{eq:jge1boltzmann} for a set of multipoles from $j=0$ to $j_{\rm max}=8$. We discretize a range of frequencies extending out to $\pm 1000$ Doppler widths from the line center, with $50$ bins per Doppler width, in the same manner as we did for the homogenous case. We assume that all the perturbed moments go to zero on the far blue side, i.e., a boundary condition of the Dirichlet type, with an additional Neumann boundary condition on the blue side for the zeroeth moment. We use a nonreflecting boundary condition at $j_{\rm max}$ to minimize the propagation of errors back to low values of $j$ \cite{Ma1995}.

Figure \ref{fig:sourceAB} shows the resulting basis solutions $\mathcal{A}_j$ and $\mathcal{B}_j$. These source terms for these solutions create regions of higher and lower optical depth, which accumulate over- and under-densities of photons in the blue damping wings of the \lya\ line. The excess photons stream between these regions, which leads to characteristic features in higher moments as well. Since there is no injection of photons, the solutions go to zero on the red-side of the line-center. 

Figure \ref{fig:sourceC} shows the solution $\mathcal{C}_j$, whose source term includes $\delta f_{\rm eq}$, which injects photons within the line. Due to these photons' large interaction cross section, local equilibrium between emission and absorption is achieved over a range of frequencies. This is reflected in the large and `truncated' peak in the monopole. Also worth noting is the characteristic double peak in the dipole, which arises due to streaming away from the central frequency.

We solve for the perturbed monopole, $\delta \overline{f}_{0}$, by averaging Eq.~\eqref{eq:basissolns} with $j=0$ over the line profile.
\begin{align}
  \delta \overline{f}_{0} & = \frac{1}{ 1 + p_{\rm sc} \overline{\mathcal{C}}_0 } \biggl[ \biggl( \delta_{\rm m} + \frac{ \delta x_{1s} }{ x_{1s} } \biggr) \overline{\mathcal{A}}_0 + \frac{\Theta}{a H} \overline{\mathcal{B}}_0 + \delta f_{\rm eq} \overline{\mathcal{C}}_0 \biggr] \mbox{.} \label{eq:monopoleavg}
\end{align}

\subsection{Perturbed recombination rate}
\label{subsec:dxedot}

\begin{figure}[t]
  \begin{center}
    \includegraphics[width=\columnwidth]{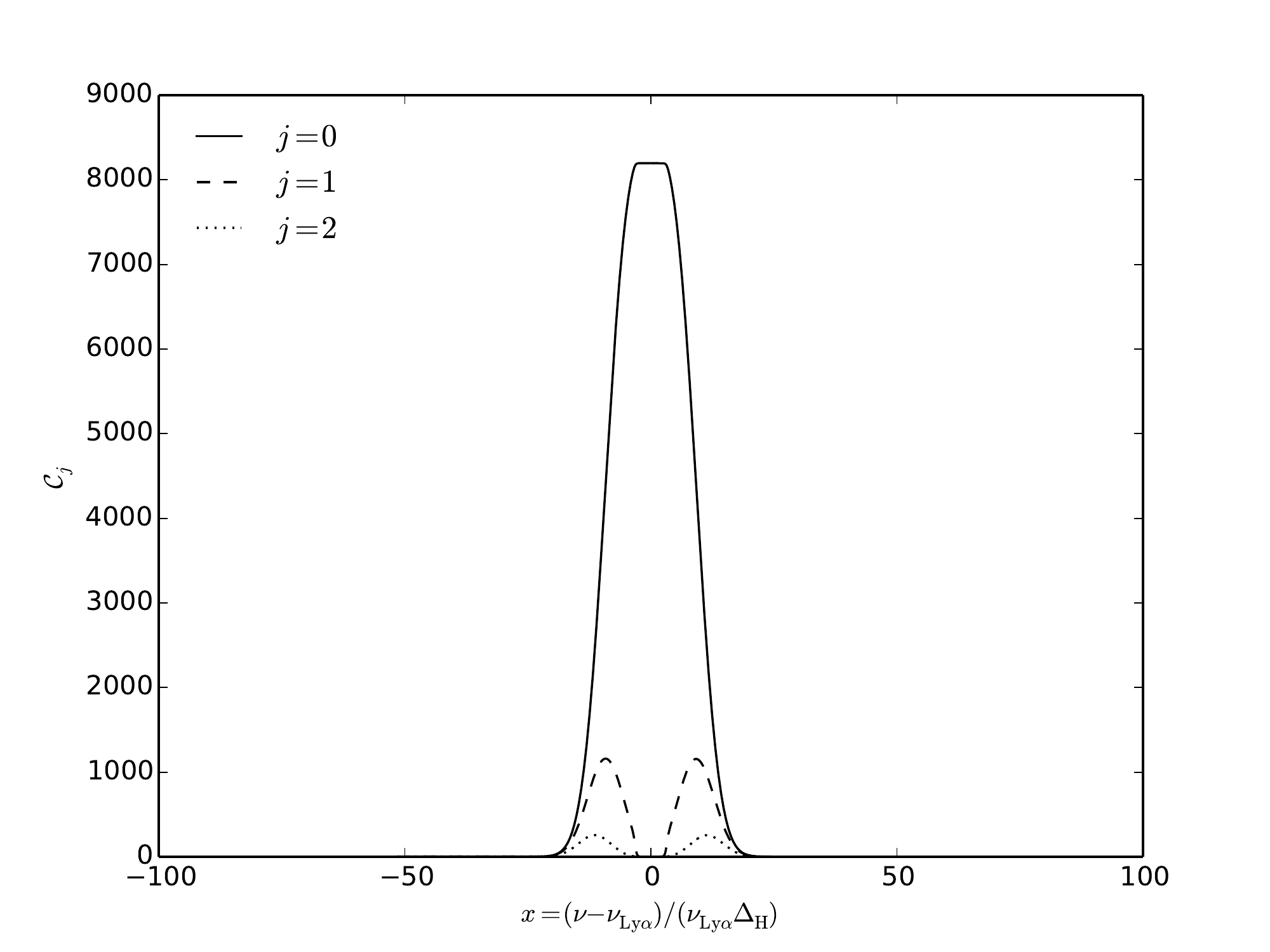}
  \end{center}
  \caption{The case of injected photons: Shown above is the solution $\mathcal{C}_j$ as defined in Eq.~\eqref{eq:basissolns}, which perturbs the equilibrium PSD. This figure is generated for $k=10^{5} ~ {\rm Mpc}^{-1}$ at redshift $z_0=1100$.}
  \label{fig:sourceC}
\end{figure}

Our goal is to compute the fluctuation in the recombination rate. We first consider the recombination rate within the \lya\ line. The linearized form of Eq.~\eqref{eq:lyarec} is
\begin{align}
  \delta \dot{x}_{1s} \vert_{\lya} = \frac{\delta x_{1s}}{x_{1s}} \dot{x}_{1s} \vert_{\lya} + 3 x_{1s} A_{\lya} \Bigl[ \delta f_{\rm eq} - \delta \overline{f}_{0} \Bigr] \mbox{.} 
\end{align}
We substitute the expression \eqref{eq:monopoleavg} for the fluctuation in the monopole averaged over the line, to write this in terms of the dimensionless solutions defined in Eq.~\eqref{eq:basissolns}.
\begin{align}
  ~~ & \!\!\!
  \delta \dot{x}_{1s} \vert_{\lya} \nonumber \\
  &= \frac{\delta x_{1s}}{x_{1s}} \dot{x}_{1s} \vert_{\lya} + 3 x_{1s} A_{\lya} \frac{1 - p_{\rm ab} \overline{\mathcal{C}}_0 }{ 1 + p_{\rm sc} \overline{\mathcal{C}}_0 } \delta f_{\rm eq} - 3 x_{1s} A_{\lya} \nonumber \\
  & ~~~\times\biggl[ \biggl( \delta_{\rm m} + \frac{ \delta x_{1s} }{ x_{1s} } \biggr) \frac{ \overline{\mathcal{A}}_0 }{ 1 + p_{\rm sc} \overline{\mathcal{C}}_0 } + \frac{\Theta}{a H} \frac{ \overline{\mathcal{B}}_0 }{ 1 + p_{\rm sc} \overline{\mathcal{C}}_0 } \biggr] \mbox{.} \label{eq:dxeLya1} 
\end{align}
Next we consider the perturbation to the two-photon decay rate from the $2s$ level. This is only sourced by changes in the level populations, since the perturbed moments of the PSD go to zero on the far red side of the line [see Figs.~\ref{fig:sourceAB} and \ref{fig:sourceC}]. The linearized form of Eq.~\eqref{eq:twophoton} is 
\begin{align}
  \delta \dot{x}_{1s} \vert_{2s} & = \frac{\delta x_{1s}}{x_{1s}} \dot{x}_{1s} \vert_{2s} + \Lambda_{2s,1s} x_{1s} \delta f_{\rm eq} \mbox{.} \label{eq:dxe2s1}
\end{align}
To close Eqs.~\eqref{eq:dxeLya1} and \eqref{eq:dxe2s1}, we need to compute the fluctuation in the equilibrium PSD, $\delta f_{\rm eq}$ (or equivalently, the population of the $n=2$ level). As in the homogenous case, we use the steady state assumption within the three level approximation, and balance the rates of the upper and lower arms of Fig.~\ref{fig:schem}. 

For the upper arm, we perturb Eq.~\eqref{eq:upperarm}, which describes the change in the population of the $n=2$ level due to photo-ionization and recombination from the continuum levels. We expect the fractional change in the population of the $n=2$ level, $x_2$, to be related to those in the other parameters of the system. The background value of $x_2$ is much smaller than the other states' populations [see discussion in Section \ref{sec:background}]. Thus, it is a good approximation to set $\delta x_e + \delta x_{1s} = 0$. Using this,
\begin{align}
  \delta \dot{x}_2 \vert_{\rm rec/ion} & = n_{\rm H} x_{\rm e}^2 \alpha_{\rm B} \Bigl[ \delta_{\rm m} + 2 \frac{\delta x_{\rm e}}{x_{\rm e}} \Bigr] - \delta x_2 \beta_{\rm B} \mbox{.} \\ 
  & = n_{\rm H} x_{\rm e}^2 \alpha_{\rm B} \delta_{\rm m}  - \Bigl[ 2 n_{\rm H} x_{\rm e} \alpha_{\rm B} + 4 f_{\rm eq} \beta_{\rm B} \Bigr]  \delta x_{1s} \nonumber \\
  & ~~~- 4 x_{\rm 1s} \beta_{\rm B} \delta f_{\rm eq} \mbox{.} \label{eq:dx2pert}
\end{align}
The rate of the lower arm is the sum of the recombination rate in the \lya\ line [Eq.~\eqref{eq:dxeLya1}] and two-photon decays from the $2s$ state [\eqref{eq:dxe2s1}]. Using Eq.~\eqref{eq:upperarm} for the background rate, and equating the sum with the RHS of Eq.~\eqref{eq:dx2pert}, we get
\begin{align}
  \delta f_{\rm eq} 
  & = 
  \Bigl[ 3 A_{\lya} \frac{1 - p_{\rm ab} \overline{\mathcal{C}}_0 }{ 1 + p_{\rm sc} \overline{\mathcal{C}}_0 } + \Lambda_{2s,1s} + 4 \beta_{\rm B} \Bigr]^{-1} \nonumber \\
  & ~~~\times \Biggl[ n_{\rm H} \frac{x_e^2}{x_{1s}} \alpha_{\rm B} \biggl( \delta_{\rm m} - \frac{\delta x_{1s}}{x_{1s}} \biggr) - 2 n_{\rm H} x_e \alpha_{\rm B} \frac{ \delta x_{1s} }{ x_{1s} } + 3 A_{\lya} \nonumber \\
    & ~~~ ~~~\times \biggl\{ \biggl( \delta_{\rm m} + \frac{ \delta x_{1s} }{ x_{1s} } \biggr) \frac{ \overline{\mathcal{A}}_0 }{ 1 + p_{\rm sc} \overline{\mathcal{C}}_0 } + \frac{\Theta}{a H} \frac{ \overline{\mathcal{B}}_0 }{ 1 + p_{\rm sc} \overline{\mathcal{C}}_0 } \biggr\} \Biggr] \mbox{.} \label{eq:feqsoln}
\end{align}

Before we compute the perturbed recombination rate, we define the quantity
\begin{align}
  \mathcal{P} & = \frac{ 3 A_{\lya} \frac{1 - p_{\rm ab} \overline{\mathcal{C}}_0 }{ 1 + p_{\rm sc} \overline{\mathcal{C}}_0 } + \Lambda_{2s,1s} }{ 3 A_{\lya} \frac{1 - p_{\rm ab} \overline{\mathcal{C}}_0 }{ 1 + p_{\rm sc} \overline{\mathcal{C}}_0 } + \Lambda_{2s,1s} + 4 \beta_{\rm B} } \mbox{.} \label{eq:deltapeebles}
\end{align}
This is the analog of Peebles' $C$ factor [see Eq.~\eqref{eq:peeblescfactor}] in the perturbed case -- it represents the probability that a fluctuation in the population of atoms in the $n=2$ level translates into one in the ground state population. 

\begin{figure}[t]
  \begin{center}
    \includegraphics[width=\columnwidth]{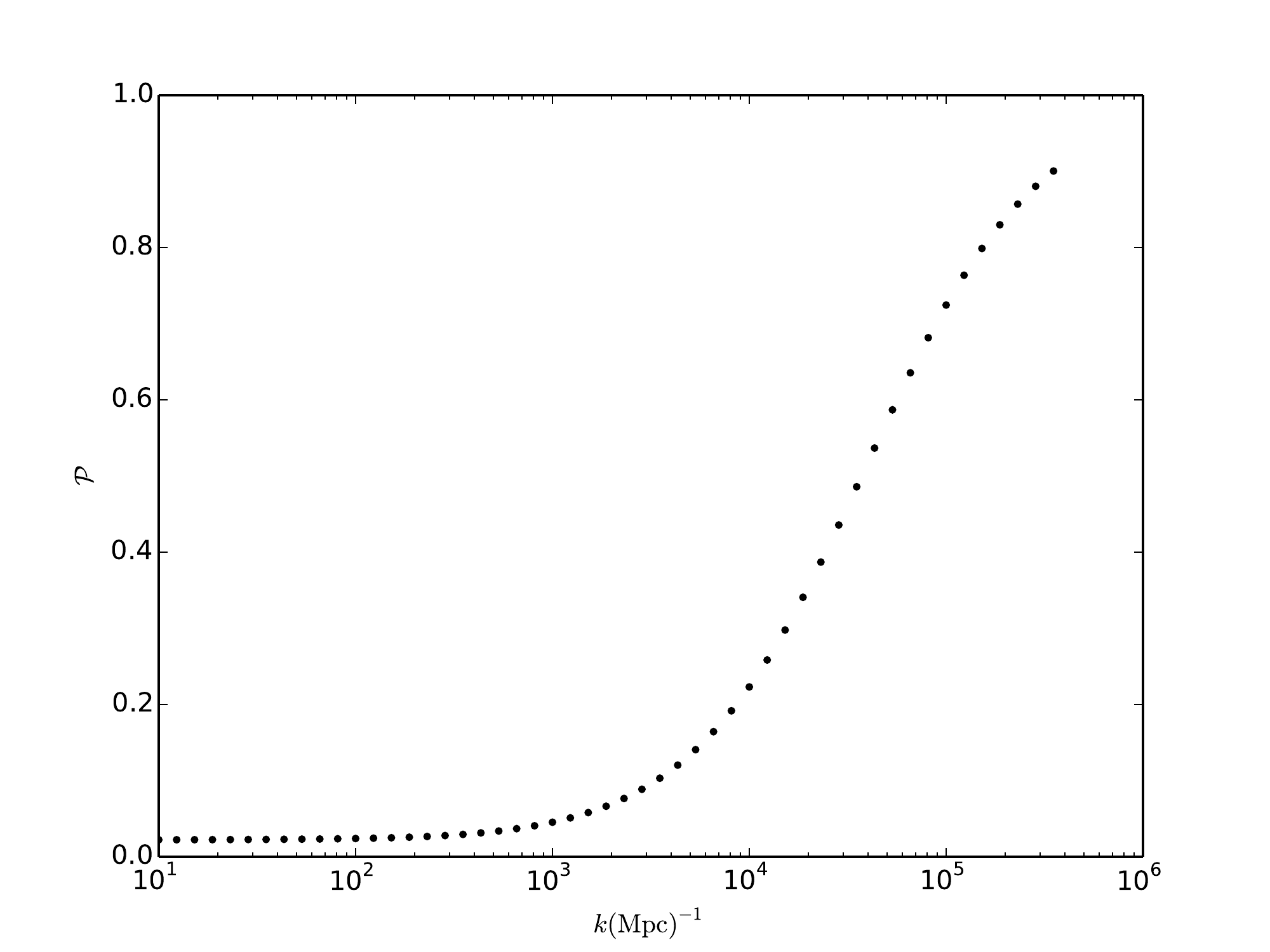}
  \end{center}
  \caption{Inhomogenous analog of the Peebles $C$ factor: The parameter $\mathcal{P}$ defined in Eq.~\eqref{eq:deltapeebles}, as function of the wavenumber, $k$. It is the probability that a fluctuation in the population of $n=2$ leads to one in that of $1s$. This figure plots values out to $k \approx 3.6 \times 10^5 {\rm Mpc}^{-1}$, up to which matter fluctuations can be assumed to be isothermal.}
  \label{fig:deltapeebles}
\end{figure}

Figure \ref{fig:deltapeebles} plots $\mathcal{P}$ as a function of the wavenumber. We observe that it asymptotes to a small value for large wavelengths. We expect this limiting value to be the Peebles $C$ factor. It approaches unity in the complementary limit of small wavelengths, but we do not show this since the assumption of the isothermal nature of such small wavelength modes breaks down at low redshifts. This turnover happens on scales of $k \approx 10^{3} \ {\rm Mpc}^{-1}$, which is large compared to the diffusion scale at line center, which was calculated in Section \ref{sec:ionizationsaha}. We give physical arguments for the large wavelength limit in Appendix \ref{sec:analytic}, and the turnover scale for small wavelengths in Appendix \ref{sec:levywalk}.

We substitute Eq.~\eqref{eq:feqsoln} into Eqs.~\eqref{eq:dxeLya1} and \eqref{eq:dxe2s1}, and use the definition of $\mathcal{P}$ to write the fluctuation in the net recombination rate as
\begin{align}
  \delta \dot{x}_{1s} \vert_{ {\lya}, 2s } 
  & = \mathcal{P} n_{\rm H} x_e^2 \alpha_{\rm B} \delta_{\rm m} + \delta x_{1s} \Bigl[ (1 - \mathcal{P}) n_{\rm H} \frac{ x_e^2 }{ x_{1s} } \alpha_{\rm B} \nonumber \\
    & ~~~ - 2 \mathcal{P} n_{\rm H} x_e \alpha_{\rm B} - 4 f_{\rm eq} \beta_{\rm B} \Bigr] - 3 ( 1 - \mathcal{P} ) x_{1s} A_{\lya} \nonumber \\
    & ~~~\times \biggl[ \biggl( \delta_{\rm m} + \frac{ \delta x_{1s} }{ x_{1s} } \biggr) \frac{ \overline{\mathcal{A}}_0 }{ 1 + p_{\rm sc} \overline{\mathcal{C}}_0 } + \frac{\Theta}{a H} \frac{ \overline{\mathcal{B}}_0 }{ 1 + p_{\rm sc} \overline{\mathcal{C}}_0 } \biggr] \mbox{.} \label{eq:perturbedxedot}
\end{align}
In Appendix \ref{subsec:diffusion_recomb}, we derive the expression for the perturbed recombination rate for wavelengths much smaller than the diffusion scale, and show that it is identical to the above expression in the limit $\mathcal{P} \rightarrow 1$.

\section{Solution for the local growth rates}
\label{sec:solution}

We solve for the local growth rates by finding the fastest growing modes of the matter field. We use Eq.~\eqref{eq:newtonianfinal} for the evolution of the matter density and velocity, and obtain the evolution equation for the perturbed ionization fraction by adding the rates of perturbed recombination due to \lya\ photons and two-photon decays from $2s$ [Eq.~\eqref{eq:perturbedxedot}] and Continuum photon transport [Eq.~\eqref{eq:xedotcont}]. We use case B recombination coefficients from \cite{Pequignot91} for numerical estimates.

\begin{figure}[t]
  \includegraphics[width=\columnwidth]{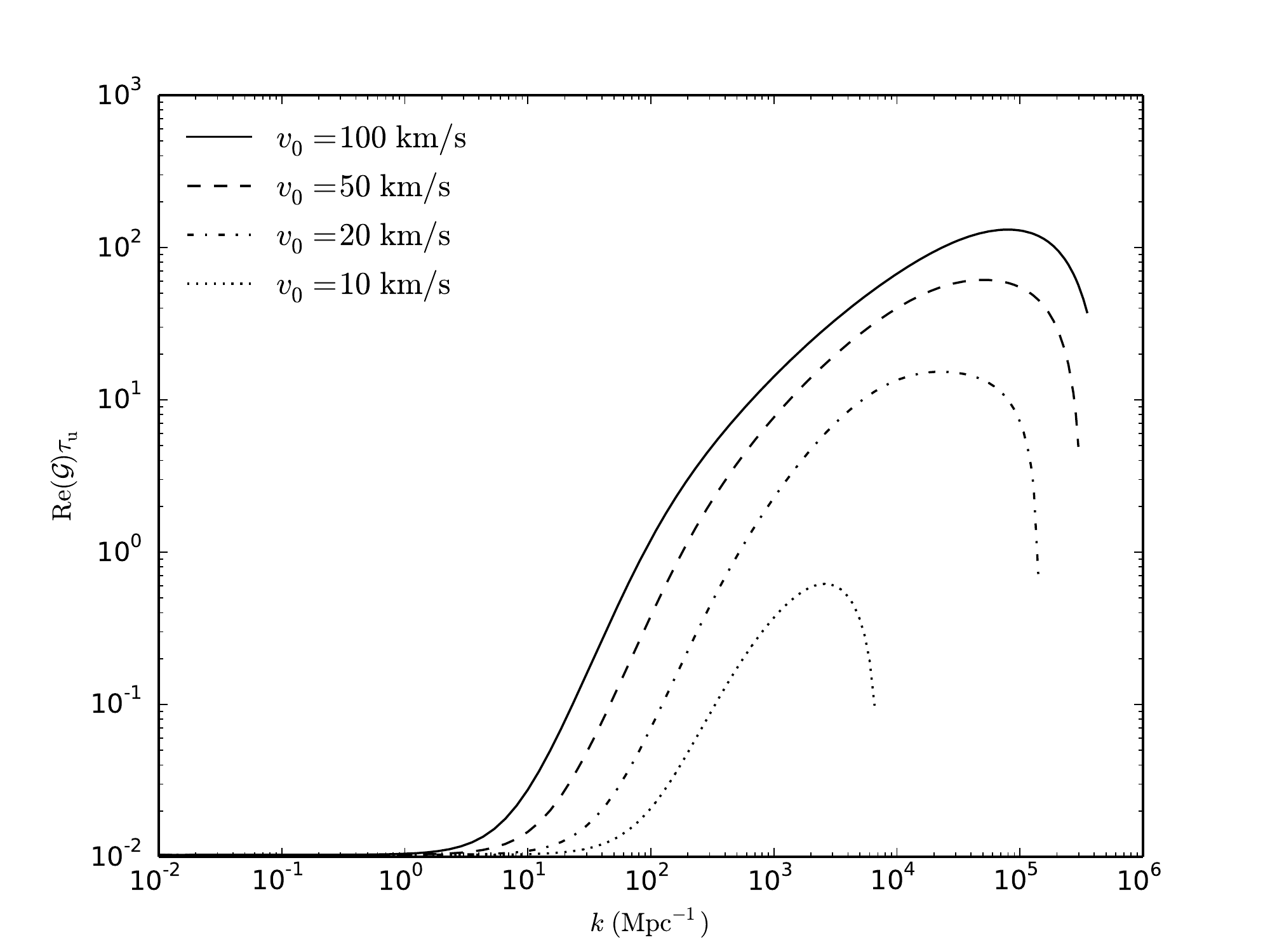}
  \caption{Maximum instantaneous growth rate for small-scale fluctuations in the matter field at recombination, normalized to the net elapsed coordinate time, $\tau_{\rm u}$. The scenario here is identical to that of Fig.~\ref{fig:growthsaha}, except that perturbed recombination is treated with full radiative transport of Continuum and Lyman-$\alpha$ photons.}
  \label{fig:growthfull}
\end{figure}

Figure \ref{fig:growthfull} plots the maximum instantaneous growth rate for small-scale matter fluctuations at recombination (normalized to the net elapsed coordinate time, $\tau_{\rm u}$ at $z=1100$) for various values of the large scale shear $v_0$. Comparison with the results of Fig.~\ref{fig:growthsaha} shows that the instability persists, and even somewhat strengthened, on intermediate scales with wavenumber $k \approx 10^2 \ {\rm Mpc}^{-1}$. However, it is cutoff on small scales due to the radiative processes described in Sections \ref{sec:continuum} and \ref{sec:Lya}. The precise wavenumber at which it is cutoff depends on the large-scale relative velocity, but is well before the saturation scale over the practically achievable range.

In the next section, we estimate the growth rates achieved due to a stochastic background relative velocity, the distribution for which was introduced in Section \ref{sec:background}. 

\section{Distribution of growth factors}
\label{sec:average}

The growth rate shown in Fig.~\ref{fig:growthfull} is a general property of the equations of motion calculated in the presence of a constant background relative velocity. In actuality, this background velocity at a given location and time is picked from the distribution of Eq.~\eqref{eq:distribution} of Section \ref{sec:background}. Moreover, values at nearby redshifts are correlated with each other. Thus we should critically consider how this distribution is sampled over time.

Towards this end, we generalize the equal-time distribution of Eq.~\eqref{eq:distribution} to
\begin{align}
  \langle v_{0,i} (\bm x, t) v^\ast_{0,j} (\bm x, t^\prime) \rangle = \frac{1}{3} \delta_{i j} \int d \ln{k} \ \mathcal{F}(k; t, t^\prime) \Delta^2_{\zeta}(k) \mbox{,} \label{eq:distribution_t} \\
  \mathcal{F}(k; t, t^\prime) = \frac{1}{k^2} [\theta_{\rm m}(k,t) - \theta_{\rm r}(k,t)] [\theta_{\rm m}(k,t^\prime) - \theta_{\rm r}(k,t^\prime)]^\ast \mbox{.}
\end{align}
The direction of the relative velocity, $\bm v_0(\bm x)$ at a given point $\bm x$, varies with time. The force term, $\bm f_{\rm rad}$, in the equation of motion \eqref{eq:forcebalance}, depends on the direction of the local wavevector relative to the background velocity. We proceed under the simplifying assumption that the fastest growing mode always aligns itself; this is true in the case where the timescale for growth is much smaller than that for change in the relative velocities. Thus the linear growth factors obtained are upper bounds to the actual ones achieved. 

\begin{figure}[t]
\includegraphics[width=\columnwidth]{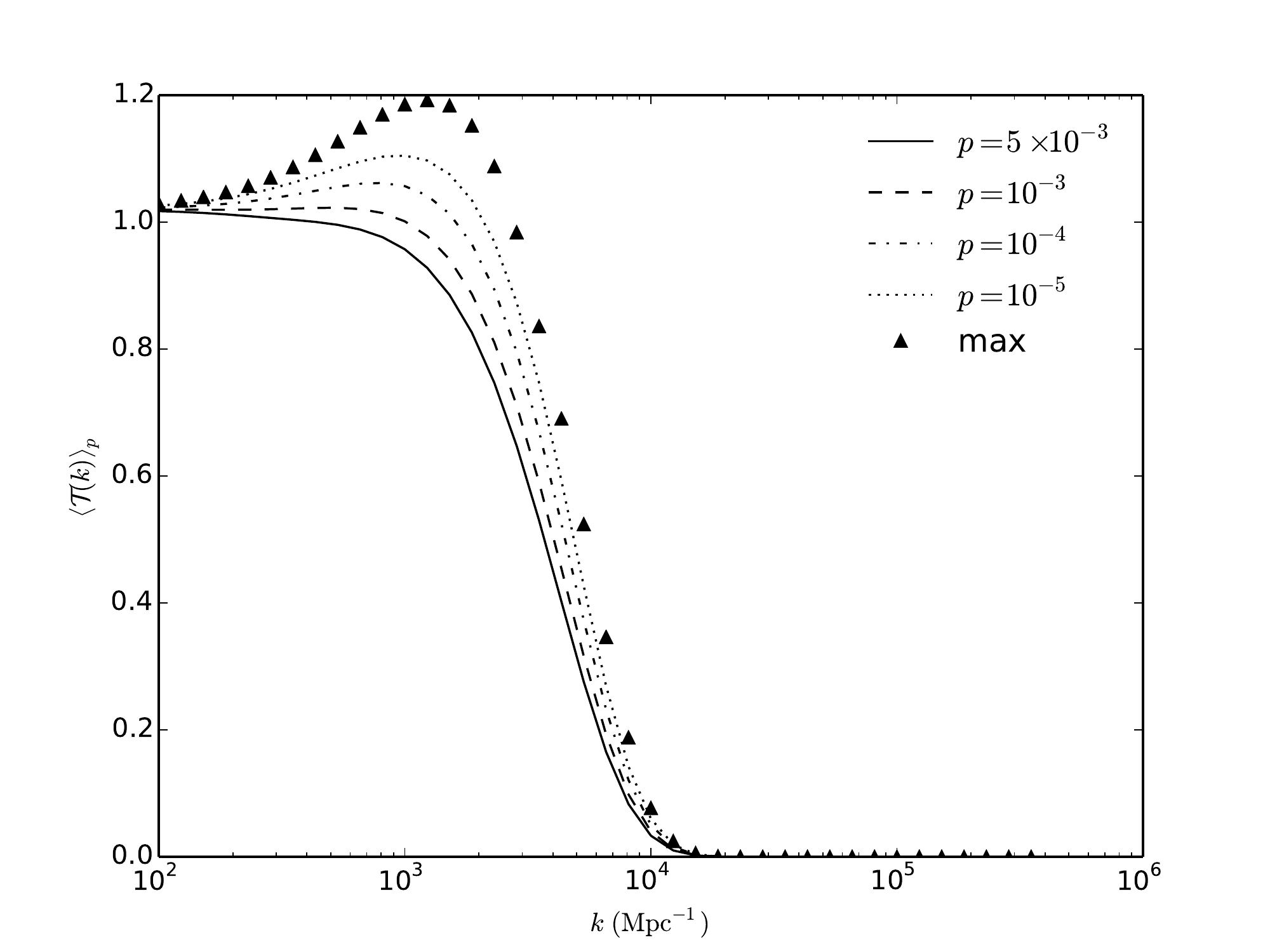}
\caption{This figure plots the mean growth factor $\left\langle \mathcal{T} (k) \right\rangle_p$ achieved in the highest $p^{\rm th}$ fraction of a sample set of $10^7$ velocity histories, for successively smaller $p$-values and a range of wavenumbers. The growth factor is normalized such that it is unity when there is no growth or suppression. Also shown is the largest growth factor for each wavenumber achieved in this sample set. Note that the growth is suppressed on scales on which the linear analysis predicts the strongest instability for large relative velocities ($k > 10^4 \ {\rm Mpc}^{-1}$, from Fig.~\ref{fig:growthfull}). }
\label{fig:growthresult}
\end{figure}
We use the notation $\mathcal{T}(\bm k, \bm x)$ to denote the net growth factor of fluctuations with wave vector $\bm k$ in a small region around a point $\bm x$. This quantity depends on the entire relative velocity history, $\bm v_0 (\bm x, t)$. At any point on the history, the growth rate is the largest eigenvalue of the equations of motion [Eqns.~\eqref{eq:deltav}, \eqref{eq:xedotlya} and \eqref{eq:xedotcont}]. As earlier, we denote this eigenvalue by $\mathcal{G}$. The growth factor in a small region around a point, $\bm x$, due to linear physics, and over the velocity history, is
\begin{equation}
  \mathcal{T} (\bm k, \bm x) = \exp{ \Bigl[ \int dt \ {\rm Re}(\mathcal{G})(\bm k, \bm v_0(\bm x,t)) \Bigr] } \mbox{.} \label{eq:maximumgrowth}
\end{equation}
Note that $\mathcal{T} (\bm k, \bm x)$ is normalized to unity in the absence of any growth or suppression. The relation in Eq.~\eqref{eq:maximumgrowth} endows the growth factor with a distribution that is inherited from that of the velocity histories. For a particular realization of the relative velocity field $\bm v_0(\bm x,t)$, the value of $\mathcal{T} (\bm k, \bm x)$ varies when both its input wave-vector $\bm k$ and position $\bm x$ are varied. However, over the entire set of realizations, there is no dependence on the direction $\hat{\bm k}$ and the position $\bm x$, due to the isotropy and homogeneity of the fluctuations underlying the relative velocities. With this understanding, we use the condensed notation $\mathcal{T}(k)$ for the growth factors. 

We generate a large number of these velocity histories in an efficient manner by sampling the distribution with the covariance matrix of Eq.~\eqref{eq:distribution_t}. We numerically sample these velocity histories at $90$ redshifts between $z=800$ and $z=1430$, and evaluate Eq.~\eqref{eq:maximumgrowth} by spline integration. In order to illustrate the tail of the growth distribution, we choose to plot the mean growth factor achieved in the highest $p^{\rm th}$ fraction of the realizations. We formally define this as 
\begin{align}
  \left\langle \mathcal{T} (k) \right\rangle_p & = \frac{1}{Np} \sum_{i = N - Np + 1}^{N} \mathcal{T}_i (k) \mbox{,} \label{eq:pvalue}
\end{align}
In this equation, $N$ is the number of realizations of the relative velocity history, $\bm v_0(\bm x, t)$, which have been sorted in increasing order of the value of $\mathcal{T}$ for the purpose of the summation. The $p$ in this definition corresponds to the usual notion of $p$-value. This use of the symbols $N$ and $p$ is restricted to this section alone, and they do not represent the number flux and momentum here.

Figure \ref{fig:growthresult} shows the tails $\left\langle \mathcal{T}(k) \right\rangle_p$ estimated from a set of $10^7$ samples of the relative velocity history, for a range of wave numbers $k$. Note that Fig.~\ref{fig:growthfull} predicts that small-scale modes of wavelengths $k \sim 10^5 \ {\rm Mpc}^{-1}$ are most unstable for a constant large-scale relative velocity. The growth factors estimated in Fig.~\ref{fig:growthfull} are optimistic for two reasons: firstly, they depend on the distribution of the {\em histories}, i.e. time-series of large-scale relative velocities, and secondly and most importantly, the instability is only active during the time where the electrons and photons are coupled, and this is much smaller than the coordinate time due to the short duration of recombination.

\section{Discussion} 
\label{sec:discussion}

The analysis in this paper accomplishes our primary goal of answering the question of the stability of small-scale fluctuations in the matter field at recombination. Our main conclusions in this regard is that {\em while growing sound wave modes exist, the amount of growth that occurs during the cosmic recombination epoch is only a fraction of an $e$-fold, and we do not expect the unstable modes to produce any phenomenological consequences}. Fluctuations with comoving wavenumbers satisfying $k > 10^2 \ {\rm Mpc}^{-1}$ are unstable in the presence of large-scale relative velocities between matter and radiation. On intermediate scales, this instability persists in the face of, and is even strengthed by the transport of continuum photons above the photo-ionization threshold, and photons within the \lya\ line of neutral hydrogen. However, this transport cuts off the growth before the saturation scale of $k \approx 10^{5} \ {\rm Mpc}^{-1}$. 

The linear analysis of the fluctuations only yields instantaneous growth rates for a constant large-scale relative velocity; the true growth factor within a given patch depends on the local relative velocity over a range of redshifts, and occurs for a duration (the width of recombination) that is shorter than the coordinate time. Accounting for this, we find no appreciable growth within a large number of random realizations of the relative velocity history. The largest growth factor achieved in our sample set, which corresponds to a $p$-value of $10^{-7}$, is slightly less than $1.2$, for modes with wavenumber $k \approx 10^{3} \ {\rm Mpc}^{-1}$. 

Along the way, we made a number of simplifying assumptions to facilitate the solution of the complicated problem of perturbed recombination. We examine a few of them below. 

The first, and most helpful one, is the three level model of the hydrogen atom, which assumes radiative equilibrium between upper levels of the true hydrogen atom. This is a good assumption at high redshifts, but becomes progressively worse as the redshift approaches $z \simeq 800$, at which point it is approximately a $10\%$ correction. In the context of homogenous recombination, there have been two approaches to deal with this -- follow the higher levels in a consistent manner \cite{HyRec}, or multiply the case-B recombination coefficient, $\alpha_{\rm B}$, with a fudge factor \cite{RecFast}. We eschew this additional complication in our preliminary analysis; instead, we generate realizations and compute growth rates only for redshifts $z \ge 800$, where the instability is expected to be strongest.

A second assumption is the equality of matter and radiation temperatures, which allows us to compute the recombination and photo-ionization rates at the CMB temperature. This is an excellent approximation for the background temperatures during the redshifts of interest due to the high Thomson scattering rates \cite{RecFast}. Its validity is much less clear in the perturbed case; a detailed discussion of timescales can be found in Ref.~\cite{Senatore2009}. In our case, the relevant comparison is the dimensionless ratio $t_{\rm sc}/t_{\rm C}$ of the sound-crossing time $t_{\rm sc} = a/(kv_{s,\rm I})$ to the Compton cooling time $t_{\rm C} = 3m_ec(1+f_{\rm He}+x_e)/(8\sigma_{\rm T}a_{\rm rad}T_\gamma^4x_e)$. These timescales are equal at a critical wavenumber $k_{\rm cr}$: sound waves are isothermal for $k\ll k_{\rm cr}$ and adiabatic (or at least decoupled from the CMB temperature) for $k\gg k_{\rm cr}$. We find that $k_{\rm cr}$ decreases with time, equaling $10^8$ Mpc$^{-1}$ at $z=1290$, $10^7$ Mpc$^{-1}$ at $z=1020$, $10^6$ Mpc$^{-1}$ at $z=870$, and $10^5$ Mpc$^{-1}$ at $z=690$. Thus for the range of redshifts we consider in this paper (up to $z=800$), we can make the isothermal approximation for modes of wavenumbers up to $k \approx 3.6 \times 10^{5} ~{\rm Mpc}^{-1}$.

Another factor we have not included in our analysis is the transport of the microwave background photons themselves between different parts of the fluctuations. Rather we have assumed that the CMB photons can freely stream through many perturbation wavelengths. At the earliest redshift considered herein, $z=1430$, the photon comoving attenuation coefficient [inverse comoving mean free path: $1/(n_{\rm H}ax_e\sigma_{\rm T})$] is 0.8 Mpc$^{-1}$. This is much smaller than the wave numbers $k$ under consideration here, justifying the treatment of the CMB as uniform.

Finally, in a larger context, this paper solves the problem of perturbed recombination for modes on very small scales. Previous work on large-scale modes relevant to the linear fluctuations in the CMB \cite{Lewis2007, LewisChallinor, Senatore2009} has shown that the ionization fraction obtained by perturbing the ODE resulting from the three-level model of the hydrogen atom is accurate enough for all practical purposes. This breaks down for very small-scale modes; modulo the proper prescription for the perturbed kinetic temperature, the method outlined in Sections \ref{sec:continuum} and \ref{sec:Lya} helps solve the problem in this limit.

\begin{acknowledgments}
  We thank Todd Thompson for bringing Shaviv's instability to our attention, and for his careful reading of an earlier draft of this paper. We also thank Cora Dvorkin for her helpful comments on the paper, and Abhilash Mishra for useful discussions. Further, we would like to acknowledge the anonymous referee for their detailed comments, which greatly improved the paper. We express our gratitude to Julien Lesgourgues and collaborators for making the CLASS code for linear perturbations freely available. During the duration of this work, TV was supported by the International Fulbright Science and Technology Award, and CH was supported by the US Department of Energy under contract DE-FG03-02-ER40701, the David and Lucile Packard Foundation, the Simons Foundation, and the Alfred P. Sloan Foundation.
\end{acknowledgments}
\appendix
\section{Lyman-$\alpha$ transport: Diffusion-dominated regime}
\label{sec:levywalk}

In this section, we study the diffusion of \lya\ photons during the epoch of recombination. In the first part of this section, we demonstrate that the length scale for their transport is much larger than the simple estimate of Eq.~\eqref{eq:naivediffusion}. In the second part, we derive a simple expression for the perturbed rate of recombination in the \lya\ and two-photon channels when the wavelength of the fluctuations is much smaller than this scale.

\subsection{Length scale for diffusion}
\label{subsec:diffusionscale}

We begin by studying the redistribution of \lya\ photons' frequency due to resonant scattering off ground-state hydrogen atoms. 

The Sobolev optical depth, $\tau_{\rm S}$, is much greater than unity at the redshift of recombination [see the estimate following Eq.~\eqref{eq:sobolev}]. The overwhelming majority of absorptions are followed by the spontaneous de-excitation of the excited atom [see Eq.~\eqref{eq:psc}]. Thus the timescale for coherent scattering is much shorter than the Hubble time for a photon in the Doppler core of the \lya\ line. A large number of scattering events effectively scrambles the initial frequency over a short time, and the emitted photon's frequency is well described by a distribution over the line profile which is incoherent with the initial one. 
\begin{align}
  p(\nu_{\rm out} \vert \nu_{\rm in}) = \phi(\nu_{\rm out}) \mbox{,} \label{eq:incoherent}
\end{align}
where we have adopted a suggestive notation for the probability distribution.

The mean free path of the scattered photon is obtained by averaging over this frequency distribution
\begin{align}
  \langle l_{\rm mfp}(\nu) \rangle = \left\langle \frac{1}{n_{1s} \sigma_{\rm sc}(\nu)} \right\rangle = \frac{1}{\tau_{\rm S} H}\frac{c}{\nu_{\lya}} \left\langle \frac{1}{\phi(\nu)} \right\rangle \rightarrow \infty \mbox{.}
\end{align}
Physically, this is a consequence of the \lya\ photon rapidly scattering out of the core into the wings, where the probability of further scattering is very small. The repeated scattering and resulting diffusion is not described by typical Brownian motion with the steps drawn from a globally Gaussian distribution. Thus the mean free path at line center, in Eq.~\eqref{eq:naivediffusion}, is a poor guide to the \lya\ transport scale.

In the rest of this section, we look at this random walk's step size distribution in more detail, and estimate a scale for the \lya\ photon transport. 

A general random walk is studied by following a collection of walkers starting at the origin. It is characterized by the distribution of their density after a given number of steps. The asymptotic form of this distribution is \cite{HughesRandWalk}
\begin{equation}
  p_N(\bm x) = \frac{1}{N^{(d/\alpha)}} L_\alpha \Bigl[ \frac{\bm x}{N^{(1/\alpha)}} \Bigr] , \qquad 0 < \alpha \leq 2 \mbox{,} \label{eq:scalingdistribution}
\end{equation}
where $d$ is the dimensionality of the random walk ($d=3$ in our case), and $L_{\alpha}[\bm x]$ is a stable distribution. Its index, $\alpha$, is fixed by the tail of the distribution of the step size: 
\begin{equation}
  \displaystyle\lim_{x \rightarrow \infty} p(x) \sim \frac{1}{x^{1 + \alpha}} \mbox{.} \label{eq:asymptote}
\end{equation}
We estimate the index in our case by marginalizing over the frequency of the scattered photon.
\begin{align}
  p(x) 
  &= \int d\nu \ p(\nu) \ p(x \vert \nu) \nonumber \\
  &\sim \int d\nu \ \phi^4(\nu) \ x^2 \exp{[- x^2 \phi^2(\nu)]} \xrightarrow[ x \rightarrow \infty ]{} x^{-2} \mbox{.}
\label{eq:probstep}
\end{align}
The argument for the scaling in Eq.~\eqref{eq:probstep} is that the dominant contribution to the integral at large step sizes, i.e., when $x \rightarrow \infty$, is from frequencies satisfying $\phi(\nu) \leq x^{-1}$; the prefactor is exponentially suppressed when we move a few Doppler widths away. Through Eq.~\eqref{eq:asymptote}, this implies a distribution of the form Eq.~\eqref{eq:scalingdistribution} for the density distribution, with an index of around unity.
 
We confirm this observation by following a large number of photons through simulated scattering events. Following each event, we redistribute the frequency incoherently according to Eq.~\eqref{eq:incoherent}, neglect any direction dependence and pick the subsequent step with a Gaussian distribution for its size, with the MFP at that frequency. 

\begin{figure}[t]
  \includegraphics[width=\columnwidth]{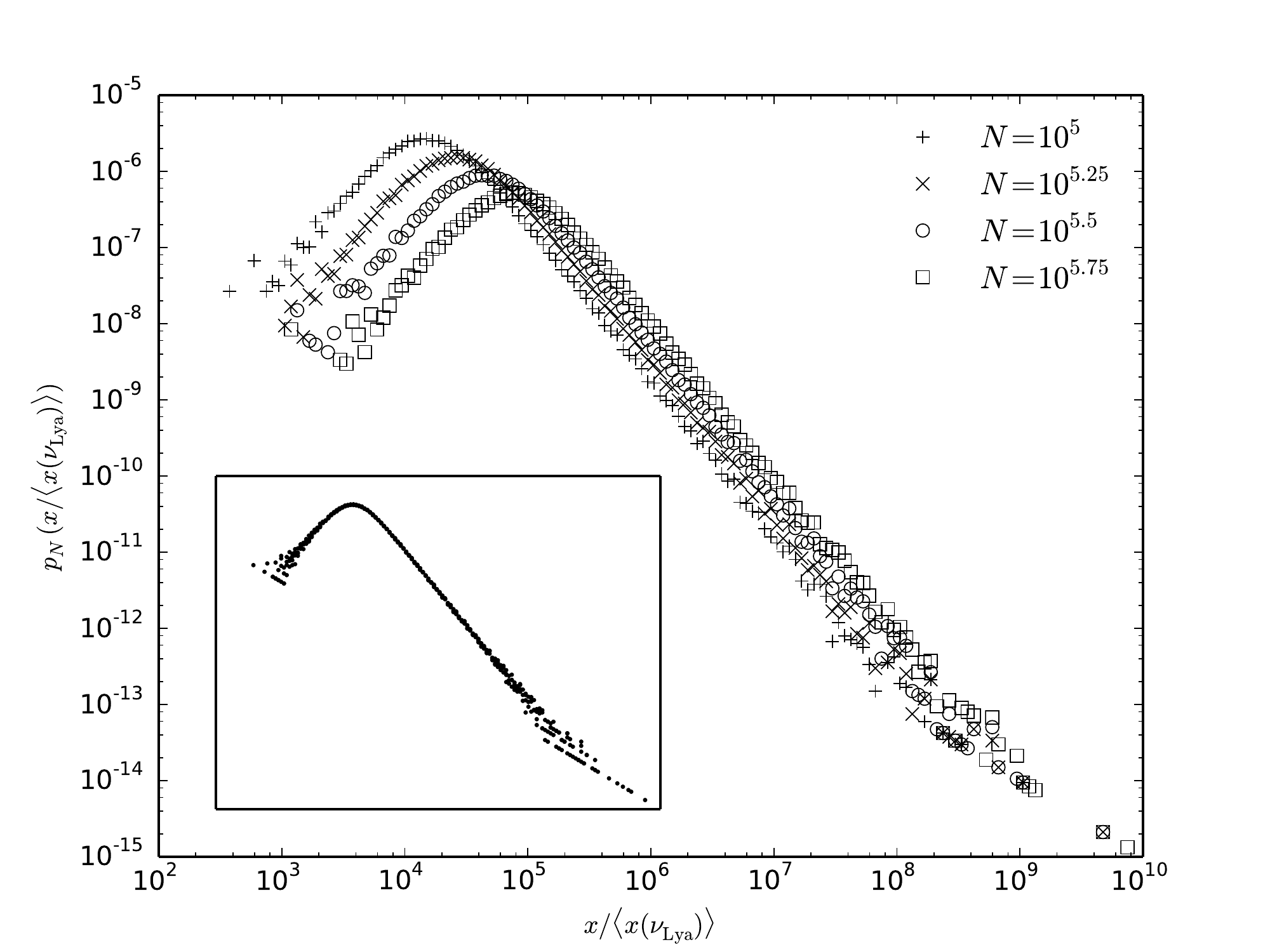}
  \caption{Histograms of the displacements of $10^5$ photons after $N$ scattering events, in units of the mean step size at the \lya\ line-center, $\langle x (\nu_{\lya}) \rangle$. They are normalized to integrate to unity. \textbf{Inset:} Demonstration of their scaling property. The histograms collapse onto a common form when the displacement, $\bm x$, is rescaled by a factor of $N^{1/\alpha}$, with the index $\alpha = 1.06$. }
  \label{fig:scalingn}
\end{figure}

Figure \ref{fig:scalingn} shows the density distributions following a large number of scattering events, $N$, and the collapse of these distributions onto a universal form when the displacements are scaled appropriately. 

The displacement does not follow the usual $\sqrt{N}$ law of Brownian motion -- instead, the histograms collapse onto a universal form when the independent variable is scaled as $N^{1/\alpha}$ with $\alpha = 1.06$. Also notable is the fact that the resulting universal form is a fat-tailed distribution which exhibits power law scaling, rather than the usual exponential falloff of the Gaussian distribution.

The quantity of direct interest for transport properties is the spread in a given time, $t$. The diverging mean-free path leads to a spread which approaches ballistic transport, hence the distributions are significantly cut off by the maximum distance $c t$.
 
As before, we directly sample the distributions through a large number of simulated scattering events. Their spread is fit by a power-law dependence of the form $\langle x^2(t) \rangle \sim t^{1.88}$.

To gauge the implications for the importance of \lya\ photon transport, we consider the various processes involved in perturbed recombination, schematically represented in Fig.~\ref{fig:schem}. The response time to a fluctuation in the ionization fraction is set by the speed of the case B recombination arm, $t_{\rm r} = (1/n_{\rm e} \alpha_{\rm B})$. From the near-ballistic transport discussed above, the time taken by a \lya\ photon to diffuse across the fluctuation is comparable to the wave crossing time $t_{\rm d} \approx (\lambda_{\rm phys}/c)$. From Fig.~\ref{fig:growthsaha}, we see that comoving wave-numbers of $k\sim 10^5 \ {\rm Mpc}^{-1}$ are most relevant for the instability. On these length scales, the wave crossing time and response time are
\begin{equation}
  t_{\rm d} \approx \frac{2\pi a}{kc} = 0.2 \ {\rm yr} \ll t_{\rm r} \approx 200 \ {\rm yr} \ {\rm at } \  z=1100 \mbox{.}
\end{equation}
These two timescales become comparable for wavenumbers $k \approx 10^2 \ {\rm Mpc}^{-1}$ at the redshift of recombination, which is when the nonlocal radiative transport starts to matter. These wavelengths are significantly larger than the simple estimate of Eq.~\eqref{eq:naivediffusion}. This is borne out by Fig.~\ref{fig:deltapeebles}. The practical consequence is that for modes with wavelengths smaller than this, perturbed recombination cannot be modeled by simply varying the cosmological parameters of the homogenous solution.

\subsection{Recombination rate in diffusion-dominated regime}
\label{subsec:diffusion_recomb}

This section uses the notation of Section \ref{sec:Lya} for the moments of the photons' phase space density. In particular, inhomogeneities in the zeroeth moment, $\delta f_{0}(\nu_{\lya}, \bm x)$, drive transport of \lya\ photons. We consider fluctuations with small enough wavelengths so that the \lya\ photons easily diffuse between the peaks and troughs. In this case, the \lya\ flux adjusts itself to wash out inhomogeneities in the zeroeth moment.
 
The population of the first excited level is set by balancing the transition rates to and from the ground state. The condition that the \lya\ phase space density is uniform yields
\begin{align}
  \delta f_{0}(\nu_{\lya}, \bm x) &= \delta \left( f_{\rm eq} \right) = 0 \mbox{,} \label{eq:perf0} \\
  \delta x_2 &= 4 f_{\rm eq} \delta x_{1s} \mbox{.} \label{eq:perx2}
\end{align}
The precise details of the radiative transfer determine the adjustment in the \lya\ flux - we avoid studying that part of the mechanism by considering the case B recombination arm of Fig \ref{fig:schem}. All that is needed to solve the recombination arm is the fluctuation in the population of the $n=2$ level, which is given by Eq.~\eqref{eq:perx2}:
\begin{align}
  \delta \dot x_{\rm e} \vert_{ {\lya}, 2s} &= \delta (- n_{\rm H} x_{\rm e}^2 \alpha_{\rm B} + x_2 \beta_{\rm B} ) \nonumber \\
  &= - n_{\rm H} x_{\rm e}^2 \alpha_{\rm B} \Bigl[ \delta_{\rm m} + 2 \frac{ \delta x_{\rm e}}{ x_{\rm e}} \Bigr] +  \beta_{\rm B} \delta x_2 \nonumber \\
  &= -n_{\rm H} x_{\rm e}^2 \alpha_{\rm B} \delta_{\rm m} -\Bigl[ 2 n_{\rm H} x_{\rm e} \alpha_{\rm B} + 4 f_{\rm eq} \beta_{\rm B} \Bigr] \delta x_{\rm e} \mbox{.}
  \label{eq:xedotlya}
\end{align}
This matches the $\mathcal{P} \rightarrow 1$ limit of the result of the complete analysis, Eq.~\eqref{eq:perturbedxedot}.

\section{Limit of weak diffusion}
\label{sec:analytic}

In this section we work out an analytical solution to the Boltzmann hierarchy in a situation with weak diffusion. This is the complementary limit to that considered in Appendix \ref{sec:levywalk}, and is realized when the wavelength of the fluctuations is much larger than the length scale for the diffusion of the \lya\ photons. We restrict ourself to the source term in Eq.~\eqref{eq:monopole} involving $\delta f_{\rm eq}$.

\subsection{Anisotropic part of hierarchy}

Let us consider the hierarchy of equations for the moments with $j \ge 1$, Eq.~\eqref{eq:jge1boltzmann}. If the range of frequencies $\Delta\nu_{\rm v}$ over which $\delta f_{j0}$ varies is larger than
\begin{equation}
\Delta\nu_{\rm mfp}
= \frac1{\tau_{\rm S}\phi(\nu)} \approx \frac{4\pi^2(\nu-\nu_{{\rm Ly}\alpha})^2}{A_{{\rm Ly}\alpha}\tau_{\rm S}}
\end{equation}
(where the approximation is in the damping wings), the photons' scattering rate is faster than that of their redshift through the frequency range of interest, and we may drop the left hand side. We expect this to be valid since $\Delta\nu_{\rm mfp}<|\nu-\nu_{{\rm Ly}\alpha}|$ in the damping wings, even out to $|\nu/\nu_{{\rm Ly}\alpha}-1|$ of order unity. 

This condition is satisfied very easily in the Doppler core due to the high scattering rates:
\begin{equation}
  \Delta\nu_{\rm mfp, core} = \nu_{\lya} \Delta_{\rm H} \frac{\sqrt{\pi}}{\tau_{\rm S}} e^{(\nu - \nu_{\lya}/\nu_{\lya} \Delta_{\rm H})^2} \ll \nu_{\lya} \Delta_{\rm H} \mbox{.}
\end{equation}
Dropping the left-hand side of Eq.~\eqref{eq:jge1boltzmann} converts the system of ODEs into an algebraic hierarchy. We can define the frequency-dependent parameter
\begin{align}
  q & = q(\nu,k) \equiv \frac{H\nu a \tau_{\rm S}\phi(\nu)}{ck} \\
  & = 1.1 \times 10^6 \left( \frac{k}{10^5 ~\rm{Mpc}^{-1} } \right)^{-1} \phi_{\rm V}(x) ~ {\rm at} ~ z = 1100 \mbox{,}
\end{align}
which is the optical depth for photons to travel a comoving distance $k^{-1}$ at that frequency. We then reduce Eq.~\eqref{eq:jge1boltzmann} to
\begin{equation}
0 = q^{-1} \left[
- \frac{j}{2j-1} \delta f_{j-1} + \frac{j+1}{2j+3} \delta f_{j+1} \right] + \delta f_{j}
\label{eq:..H2}
\end{equation}
(for $j\ge 1$). It is convenient at this point to transform back to angle-space, i.e. to work with the function $\delta f(\nu,k,\mu)$. Multiplying Eq.~\eqref{eq:..H2} by $2$ and using the inverse transformation of Eq.~\eqref{eq:moments}, we see that
\begin{align}
  0 &= \int_{-1}^1 d\mu \ \delta f(\mu) \Bigl[ 
    i q^{-1} j P_{j-1}(\mu)
    + i q^{-1} (j+1) P_{j+1}(\mu)
    \nonumber \\ 
    & ~~~+ (2j+1) P_j(\mu) \Bigr] \mbox{.}
\end{align}
Using the multiplication formula for the Legendre polynomials gives
\begin{equation}
  0 = (2j+1) \int_{-1}^1 d\mu \ P_j(\mu) \ \delta f(\mu) ( iq^{-1} \mu + 1 ) .
\end{equation}
This holds for all $j\ge 1$, hence the solution is that the combination $(i q^{-1} \mu + 1) \delta f(\mu)$ must be a constant independent of $\mu$:
\begin{equation}
  \delta f(\mu) = \mathcal{F} \frac{1}{1 + i q^{-1} \mu}.
\end{equation}
In particular, the relation between the first and zeroeth moments is
\begin{align}
  \frac{\delta f_{1}(\nu,k)}{\delta f_{0}(\nu,k)} & = 3 i \frac{ \int_{-1}^1 d\mu \ \mu \ \delta f(\nu,k,\mu) }{ \int_{-1}^1 d\mu \ \delta f(\nu,k,\mu) }
\nonumber \\
& = 3 i \frac{ \int_{-1}^1 d\mu \ \mu \ (1 + i q^{-1} \mu)^{-1} }{ \int_{-1}^1 d\mu \ (1 + i q^{-1} \mu)^{-1} }
\nonumber \\
& = -3 q \Bigl[ 1 - \frac{1/q}{\arctan{(1/q)}} \Bigr] \mbox{.}
\label{eq:..ratio}
\end{align}

\subsection{The isotropic part}

It remains to solve the equation for $\delta f_{0}(\nu,k)$. We substitute the relation \eqref{eq:..ratio} into Eq.~\eqref{eq:monopole}, and retain the source term of interest to get the Boltzmann equation for this moment
\begin{align}
  ~~ & \!\!\!
  \frac{\partial \delta f_{0}}{ \partial \nu} \nonumber \\
  & = - \tau_{\rm S} \phi(\nu) \left[ \delta f_{\rm eq} - p_{\rm sc} \delta \overline{f}_{0} \right] - \tau_{\rm sc} \frac{\nu_{\lya}^2 \Delta_{\rm H}^2}{2} \frac{\partial}{\partial \nu} \Bigl[ \phi(\nu) \frac{\partial \delta f_0}{\partial \nu} \Bigr] \nonumber \\
  & ~~~+ \tau_{\rm S} \phi(\nu) \Bigl[ p_{\rm ab} - \Bigl\{ 1 - \frac{1/q}{\arctan{(1/q)}} \Bigr\} \Bigr] \delta f_0 \mbox{.} \label{eq:..H3}
\end{align}
The boundary condition is that $\delta f_{00,+} = 0$ (i.e. no perturbation to the incoming radiation on the blue side of the line). The solution $\mathcal{C}_0(\nu)$ of Eq.~\eqref{eq:basissolns} is determined by setting $\delta f_{\rm eq}-p_{\rm sc}\delta\bar f_{00}=1$ in Eq.~(\ref{eq:..H3}). 

We examine the simplest case, where the frequency diffusion term is negligible. In the limit we are considering in this section, the wave-number $k \rightarrow 0$. In that case, the parameter $q \rightarrow \infty$, and the term in curly braces on the RHS of Eq.~\eqref{eq:..H3} approaches zero. Taking this limit, we have
\begin{equation}
  \frac{\partial\delta f_{0}}{\partial\nu} = \tau_{\rm S} \phi(\nu) p_{\rm ab} \delta f_{0} - \tau_{\rm S}\phi(\nu).
\end{equation}
Defining the cumulative distribution function of the profile ${\cal X} = \int d\nu \ \phi(\nu)$ (so that ${\cal X}$ ranges from $0$ at the red side of the line to $1$ at the blue side), we may solve this equation to yield
\begin{equation}
\delta f_{0}(\nu) = \frac1{p_{\rm ab}} \left[ 1 - e^{p_{\rm ab}\tau_{\rm S}({\cal X}-1)} \right].
\end{equation}
Averaging over the line profile is equivalent to the integration $\int_0^1 d{\cal X}$:
\begin{equation}
  \overline{\mathcal{C}}_0 = \delta \overline{f}_{0} = \frac1{p_{\rm ab}} \left( 1 - \frac{1 - e^{-\tau_{\rm S}p_{\rm ab}}}{\tau_{\rm S}p_{\rm ab}} \right).
\end{equation}
It follows that
\begin{equation}
  \frac{1-p_{\rm ab}\overline{\mathcal{C}}_0}{1+p_{\rm sc}\overline{\mathcal{C}}_0}
= \frac1{\tau_{\rm S}}
\frac{1-e^{-\tau_{\rm S}p_{\rm ab}}}{
1 - p_{\rm sc}(1 - e^{-\tau_{\rm S}p_{\rm ab}})/(\tau_{\rm S}p_{\rm ab}) }.
\end{equation}
In the relevant optically thick limit of $\tau_{\rm S}p_{\rm ab}\gg 1$, this becomes equivalent to the usual Sobolev escape probability, $\approx 1/\tau_{\rm S}$. Substitution into the definition of $\mathcal{P}$ in Eq.~\eqref{eq:deltapeebles} recovers the standard Peebles' $C$ factor of Eq.~\eqref{eq:peeblescfactor}.

\bibliography{references_new}
\end{document}